\documentclass[fleqn,10pt]{SelfArx} 
\usepackage[numbers,sort&compress]{natbib}
\usepackage[font={small,singlespacing}]{caption}
\usepackage{setspace}
\usepackage{hyperref}
\usepackage{lettrine}
\usepackage{booktabs} 
\usepackage{array} 

\definecolor{color1}{RGB}{39,45,102} 
\definecolor{color2}{RGB}{240,240,240} 

\newlength{\tocsep} 
\JournalInfo{Nov 2014} 
\Archive{} 

\PaperTitle{Social media fingerprints of unemployment}%

\Authors{Alejandro Llorente\textsuperscript{1,2}, Manuel Garc\'{\i}a-Herranz\textsuperscript{3}, Manuel Cebrian\textsuperscript{4,5}, Esteban Moro \textsuperscript{1,2}*}
\affiliation{\textsuperscript{1}\textit{Instituto de Ingenier\'{\i}a del Conocimiento, Universidad Aut\'onoma de Madrid, Madrid 28049, Spain}}
\affiliation{\textsuperscript{2}\textit{Departamento de Matem\'aticas \& GISC, Universidad Carlos III de Madrid, Legan\'es 28911, Spain}}
\affiliation{\textsuperscript{3}\textit{UNICEF Innovation Unit,  New York, NY 10017, USA}}
\affiliation{\textsuperscript{4}\textit{Department of Computer Science and Engineering, University of California at San Diego, La Jolla, CA 92093, USA}}
\affiliation{\textsuperscript{5}\textit{National Information and Communications Technology Australia, Melbourne, Victoria 3003, Australia}}
\affiliation{*\textbf{Corresponding author}: emoro@math.uc3m.es} 

\newcommand{\keywordname}{Keywords} 

\Abstract{ Recent wide-spread adoption of electronic and pervasive technologies has enabled the study of human behavior at an unprecedented level, uncovering universal patterns underlying human activity, mobility, and inter-personal communication. In the present work, we investigate whether deviations from these universal patterns may reveal information about the socio-economical status of geographical regions.
We quantify the extent to which deviations in diurnal rhythm, mobility patterns, and communication styles across regions relate to their unemployment incidence. For this we examine a country-scale publicly articulated social media dataset, where we quantify individual behavioral features from over 145 million geo-located messages distributed among more than 340 different Spanish economic regions, inferred by computing communities of cohesive mobility fluxes. We find that regions exhibiting more diverse mobility fluxes, earlier diurnal rhythms, and more correct grammatical styles display lower unemployment rates. As a result, we provide a simple model able to produce accurate, easily interpretable reconstruction of regional unemployment incidence from their social-media digital fingerprints alone. Our results show that cost-effective economical indicators can be built based on publicly-available social media datasets.}

\begin{document}

\flushbottom 

\maketitle 


\thispagestyle{empty} 

\setstretch{0.92}

\lettrine[nindent=0pt, lines=2]{H}{uman} behavior is closely intertwined with socioeconomical status, as many of our daily routines are driven by activities related to maintain, to improve, or afforded by such status~\cite{becker1976economic,granovetter1985economic,
camerer2011advances}. From our movements around the city, to our daily schedules, to the communication with others, humans perform different actions along the day that reflect and impact their economical situation. The distribution of different individual behaviors across neighborhoods, municipalities, or cities impacts the economical development of those geographical areas, and in turn to that of the whole country~\cite{ glaeser1991growth,bettencourt2007growth, batty2008size,milgram1974experience,pan2013urban,gonzalez2008understanding}. 
Detecting patterns and quantifying relevant metrics to unveil the complex relationship between geography and collective behavior is thus of paramount importance for understanding the economical heart-beat of cities, and the structure of inter-city networks, and thus to economic planning, educational policy, urban planning, transportation design, and other large-scale societal problems~\cite{Calabrese2013301,cheng:footprints,Cho:2011:FMU:2020408.2020579,sunlijun,Eagle21052010}.

Much knowledge about how mobility, social communication and education affect the economical development of cities has been being obtained through complex and costly surveys, with an update rate ranging from fortnights (unemployment) to decades (census)~\cite{henrich2001search,krieger1997measuring,groves2013survey}. At the same time, the recent availability of vast and rich datasets of individual digital fingerprints has increased the scale and granularity at which we can measure these behavioral features, reduced the cost and update rate of these measurements, and provided new opportunities to combine them with more traditional socio-economical surveys~\cite{lazer2009life,Eagle21052010,smith2013finger,Soto:2011:PSL:2021855.2021893,preis2012quantifying,antenucci2014using}. 

In this work we provide a proof of concept for the use of social media individual digital fingerprints to infer city-level behavioral measures, and then uncover their relationship with socioeconomic output. We present a comprehensive study of the different behavioral traces that can be extracted from social media: (i) technology adoption from (social media) user demographics, (ii) mobility patterns from geo-located messages, (iii) communication patterns from exchanged messages, and (iv) content analysis from the published messages. 
To this end, we use a country-scale publicly articulated social media dataset in Spain, where we
infer behavioral patterns from almost 146 million geo-located messages. We match this dataset with the granular
unemployment at the level of municipality, measured at the peak of the Spanish financial crisis (2012--2013). We consider unemployment to be the most important signal for the socioeconomic status of a region, since the effects of the crisis have had a very large impact in terms of unemployment in the country (around 9.2\% in 2005, more than 26\% in 2013). 

Our extensive investigation of this large variety of traces in a large social media dataset allows us not only to build an accurate model of unemployment impact across geographical areas, 
but also to compare globally previously reported metrics in diverse works and datasets, as well as asses 
their relevance and uniqueness to understand economical development \cite{Eagle21052010,antenucci2014using,smith2013finger,Soto:2011:PSL:2021855.2021893,hawelka2013geo,lathia2012hidden,frias2010socio,gutierrez2013evaluating,smith2013ubiquitous}. As we will show, technology adoption, mobility, diurnal activity, and communication style metrics carry a different weight in explaining unemployment in different geographical areas. Our goal is not to state causality between unemployment and the extracted metrics but to uncover the relationship emerging when we observe the economical metrics of cities and the social behavior at the same time.

\section{Social media dataset and functional partition of cities}

Twitter is a microblogging online application where users can express their opinions, share content and receive information from other users in text messages of 140 characters long, commonly known as {\it tweets}. Users can interact with other users by mentioning them or retweeting (share someone's tweet with your followers) their content. Some of these tweets contain information about the geographical location where the user was located when the tweet was published; we refer to them as geo-located tweets.
\begin{figure*}[!ht]
\centering
	\includegraphics[width=0.8\textwidth]{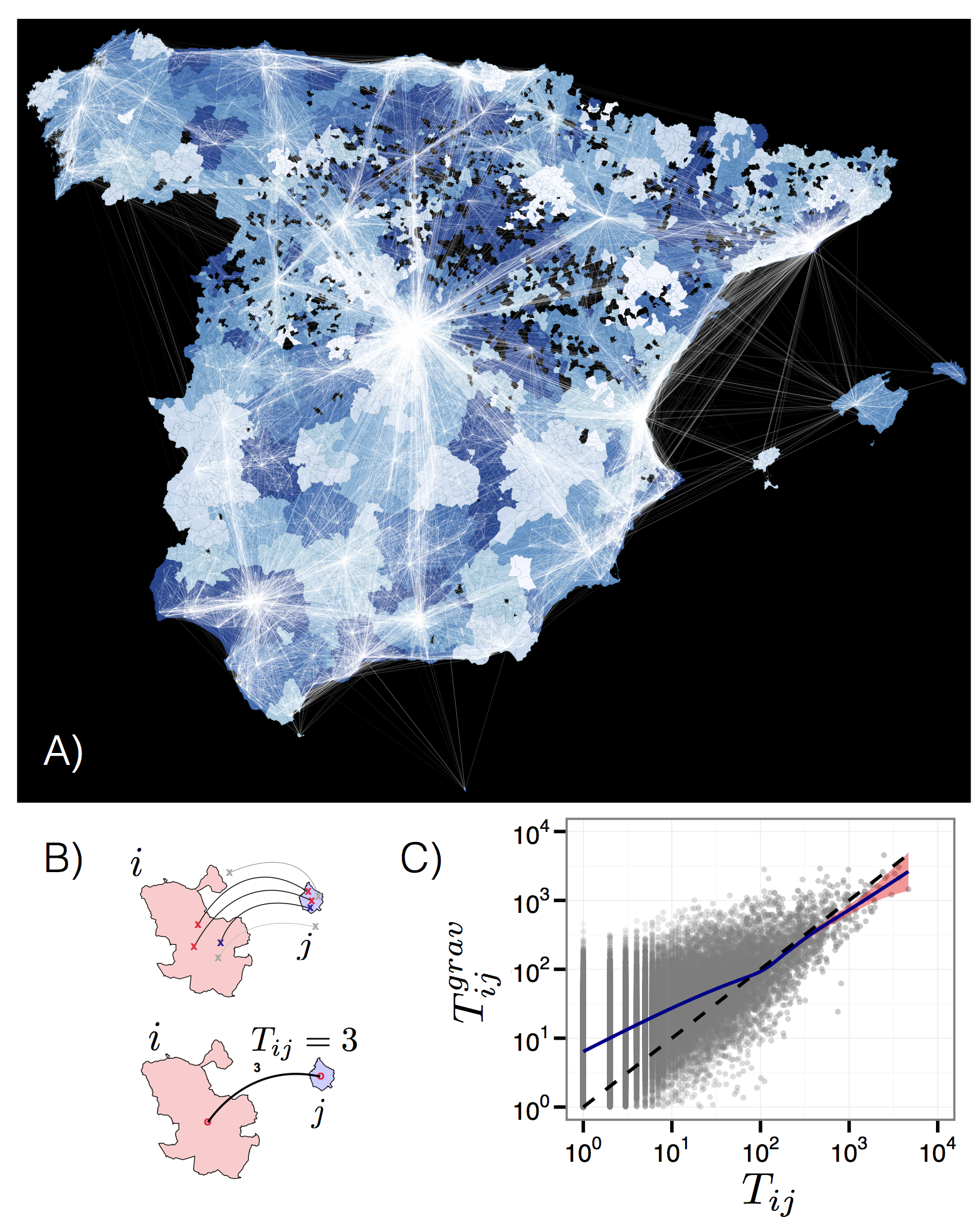}
	\caption{A) Map of the mobility fluxes $T_{ij}$ between municipalities based on Twitter inferred trips (white). Infomap communities detected on the network $T_{ij}$ are colored under the mobility fluxes (blue colors). B) Mobility fluxes $T_{ij}$ between municipalities $i$ and $j$ are constructed by aggregating the number of trips between them. C) Correspondence between the observed fluxes $T_{ij}$ and the fitted gravity model fluxes. Dashed line is the $T_{ij} = T_{ij}^\mathrm{grav}$ while the (blue) solid line is an conditional average of $T_{ij}^\mathrm{grav}$ for fixed values of $T_{ij}$.}
	\label{fig:mobility_graph}
\end{figure*}

To perform our analysis, we consider 19.6 million geo-located Twitter messages (tweets), collected through the public API provided by Twitter from continental Spain, ranging from 29th November 2012 to 30th June 2013. Tweets were posted by (properly anonymized) 0.57 Million unique users and geo-positioned in 7683 different municipalities. We observed a large correlation (Pearson's coefficient $\rho = 0.951 [0.949,0.953]$) between the number of geopositioned tweets per municipality and the municipality's population. On average we find around 50 tweets per month and per 1000 persons in each municipality. 

Despite this high level of social media activity within municipalities, we find their official administrative areas not suitable to study socio-economical activity: administrative boundaries between municipalities reflect political and historical decisions, while economical trade and activity often happens across those boundaries. The result is that municipalities in Spain are artificially diverse, ranging from a municipality with only 7 inhabitants to other with population 3.2 million. 
Although there exists natural aggregations of municipalities in provinces (regions) or statistical/metropolitan areas (NUTS areas), we have used our own procedure to detect economical areas. In particular, we have used user daily trips between pairs of municipalities as a measure of the economic relatedness between said municipalities. We say that there is a daily trip between municipality $i$ and $j$ if a user has tweeted in place $i$ and $j$ consecutively within the same day. In our dataset we find 1.9 million trips by 0.22 million users. 
With those trips we construct the daily mobility flux network $T_{ij}$ between municipalities as the number of trips between place $i$ and $j$ (see \ref{fig:mobility_graph}B). Remarkably, the statistical properties of trips and of the mobility matrix $T_{ij}$ coincide with those of other mobility datasets (see SI section 2): for example, trip distance $r$ and elapsed time 
$\delta t$ are power-law distributed with exponents $P(r) \sim r^{-1.67}$ and $P(\delta t) \sim \delta t^{-0.62}$, very similar to those found in the literature \cite{gonzalez2008understanding,hawelka2013geo}. And the mobility fluxes $T_{ij}$ are well described by the Gravity Law ($R^2 = 0.80$) 
\cite{erlander1990gravity}

\begin{equation}\label{eq:gravity}
T_{ij} \simeq T_{ij}^{\mathrm{grav}} = \frac{P_i^{\alpha_i} P_j^{\alpha_j}}{d_{ij}^\beta}
\end{equation}

where $P_i$ and $P_j$ are the populations of municipalities $i$ and $j$ and $d_{ij}$ is the distance between them. Similarly, the exponents in (\ref{eq:gravity}) are very similar to those reported in other works $\alpha_i \simeq \alpha_j = 0.48$ and $\beta \simeq 1.05$ \cite{simini2012universal,hawelka2013geo}. These results suggest that detected mobility from geo-located tweets is a good proxy of human mobility within and between municipalities \cite{lenormand2014cross}.

We use the network of daily fluxes between municipalities $T_{ij}$ to detect the geographical communities of economical activity. To this end we employ standard partition techniques of the mobility network $T_{ij}$ using graph community finding algorithms. This technique has been applied extensively, specially with mobile phone data, to unveil the effective maps of countries based on mobility and/or social interactions of people\cite{Barthelemy20111,expert2011uncovering,sobolevsky2013delineating}. In our case, we have used the Infomap algorithm \cite{rosvall2008maps} and found 340 different communities within Spain. For further details about the comparison among different state-of-art community detection algorithms executed on the inter-city graph, see SI section 3. The average number of municipalities per community is 21, and the largest community contains 142 municipalities. The communities detected have very interesting features (see SI section 3): (i) they are cohesive geographically (see figure \ref{fig:mobility_graph}), (ii) they are statistically robust against randomly removal of trips in our database (SI table S2), and (iii)  modularity of the partition is very high ($~ 0.76$, see SI table S3). Finally, (iv) the partition found has some overlap (77\% of Normalized Mutual Information, NMI, see \cite{danon2005comparing}) with coarser administrative boundaries like provinces (regions) (see SI section 3 for details). But interestingly, it shows a larger overlap (83\% of NMI) with {\em comarcas} (counties), areas in Spain that reflect geographical and economical relations between municipalities . This result shows that the mobility detected from geo-located tweets and the communities obtained are a good description of economical areas. 

In the rest of the paper, we restrict our analysis to the geographical areas defined by the Infomap detected communities (see figure \ref{fig:mobility_graph}). For statistical reasons, we discard communities which are not formed by at least 5 municipalities. Despite this sampling, 96\% of the total country population is considered in our analysis. Our results in the rest of the paper also hold for municipalities, counties or provinces, though with lower statistical power (see SI section 9).


\section{Social media behavioral fingerprints}
The goal of this work is to quantify how and what behavioral features can be extracted from social media and then related back to the to the economical level of cities. To this end, we define four groups of measures that have been widely explored in other fields like economy or social sciences. These four types measures rely on the identification of the place where users live. Instead of using information in the user profile, we analyze the places where the user has tweeted and we set as {\it hometown} of the user the municipality where he/she has tweeted with the highest frequency, a method usually employed in mobile phone and social media \cite{cheng:footprints,hawelka2013geo}. To this end we select those users with more than 5 geo-located tweets in our period and which have tweeted at least 40\% of their tweets in a given municipality, which we will consider their hometown. After this filtering we end up with 0.32 million users and we can then define the twitter population $\pi_i$ in area $i$ as the number of users with their hometown within area $i$. We obtain a very high correlation between $\pi_i$ and population of the cities $P_i$ in the national census $\rho = 0.977 [0.976, 0.978]$ which provides an indirect validation of our approach with the present data. However not all demographic groups are equally represented in the our twitter database. As shown in the SI section 4, Twitter user demographics in Spain obtained from surveys \cite{adigital2012} show that age groups above 44 years old are under-represented. Thus our results would mainly describe the socio-economical status of people below 44 years old. Employment analysis is then performed in different age groups: unemployment for people below 25 years old, between 25 and 44 years old and older than 44 years old. Finally, we have chosen the unemployment reported officially at the end of our observation time window (June 2013), but our results are not affected by the month selected, see SI Section 7. 
\begin{figure*}[!ht]
\centering
	\includegraphics[width=0.85\textwidth]{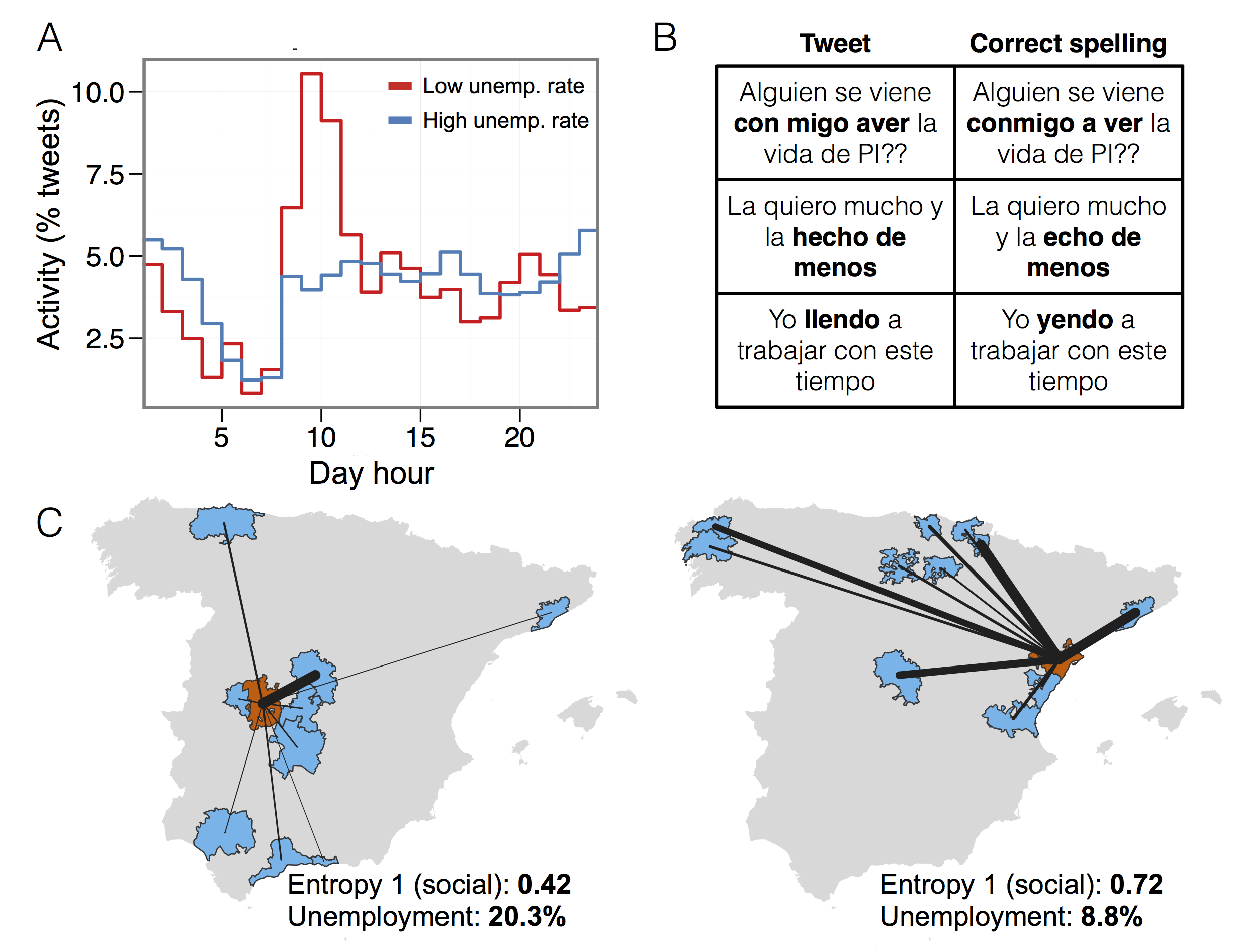}
	\caption{Examples of different behaviour in the observed variables and the unemployment. In A, we observe that two cities with different unemployment levels have different temporal activity patterns. Figure C show how communities (red) with distinct entropy levels of social communication with other communities (blue) may hold different unemployment intensity: left map shows a highly focused communication pattern (low entropy) while right map correspond to a community with a diverse communication pattern (high entropy). Finally, figure B shows some examples of detected misspellings in our database using 618 incorrect expressions (see SI Section 6) such as ``Con migo'', ``Aver'' or ``llendo''.\label{fig:variables_explanation}}
\end{figure*}

\begin{figure*}[!ht]
\centering
	\includegraphics[width=\textwidth]{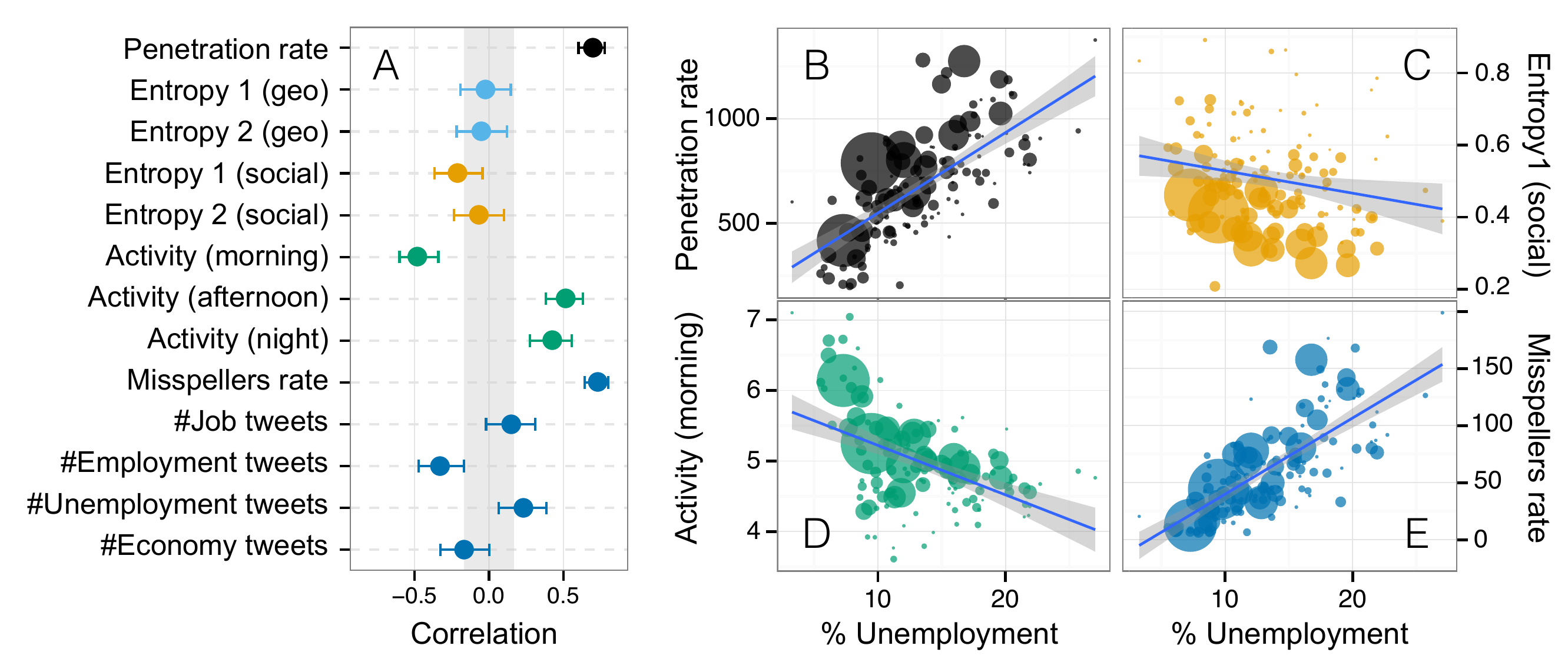}
	\caption{A) Correlation coefficient of all the extracted Twitter metrics grouped by technology adoption (black) geographical diversity (orange), social diversity (light blue), temporal activity (green) and content analysis (dark blue). Error bars correspond to 95\% confidence intervals of the correlation coefficient. Gray area correspond the statistical significance thresholds. Panels  B, C, D and E show the values of 4 selected variables in each geographical community against its percentage of unemployment. Size of the points is proportional to the population in each geographical community. Solid lines correspond to linear fits to the data.}
	\label{fig:corr_unemployment}
\end{figure*}

For every considered region, we investigate the officially reported unemployment for different age groups and a number of metrics related to social media activity. Some of those metrics are already reported in the literature, but some others are introduced in this work. Specifically we consider:
\begin{itemize} 
\item {\it Social media technology adoption}: we can use twitter penetration rate $\tau_i = \pi_i / P_i$ in each area $i$ as a proxy of technology adoption. Recent works have shown that indeed there is a correlation between country GDP and twitter penetration: specifically, it was found that a positive correlation between $\tau_i$ and GDP at the country level \cite{hawelka2013geo}. However, in our data we find the opposite correlation (see figure \ref{fig:corr_unemployment}), namely, that the larger the penetration rate the bigger the unemployment is, which suggest that the impact of technology adoption at country scale is different of what happens within an (industrialized) country where technology to access social media is commoditized. 
\item {\it Social media activity}: regions with very different economical situations should exhibit different patterns of activity during the day. Since working, leisure, family, shopping, etc. activities happen at different times of the day, we might observe different daily patterns in regions with different socio-economical status. For example, we hypothesize that communities with low levels of unemployment will tend to have higher activity levels at the beginning of a typical weekday. This is indeed what we find: figure \ref{fig:variables_explanation}A shows the hourly fraction of tweets during workdays of two communities with very different rate of unemployment. As we can observe, both profiles are quite different and, in the case of low unemployment, we find a strong peak of activity between 8 and 11am (morning), and lower periods of activity during the afternoons and nights. We encode this finding in $\nu_{\mathrm{mrng},i}$, $\nu_{\mathrm{aftn},i}$, and $\nu_{\mathrm{ngt},i}$ the total fraction of tweets happening in geographical area $i$ between 8am and 10am, 3pm and 5pm, and 12am and 3am respectively. Figure \ref{fig:corr_unemployment} shows a strong negative correlation between $\nu_{\mathrm{mrng},i}$ and the unemployment for the communities in our database and positive correlation with $\nu_{\mathrm{aftn},i}$, and $\nu_{\mathrm{ngt},i}$.
	\item {\it Social media content}: some works have observed a correlation between the frequency of words related to work conditions \cite{antenucci2014using} or looking forward thinking searches \cite{preis2012quantifying} to the economical situation of countries. In our case we also find that there is a moderate positive correlation between the fraction of tweets $\mu_i$ mentioning {\em job} or {\em unemployment} terms and the observed unemployment, while the correlation is negative for the number mentions to {\em employment} or the {\em economy}. However, we have tried a different approach by measuring the relation between the way of writing and the educational level \cite{davenport2014readability}. To this end, we build a list of 618 misspelled Spanish expressions and extract the tweets of the dataset containing at least one of these words (see SI section 6 for further details about how these expressions were collected). We only consider tweets in Spanish, detected with a N-grams based algorithm. Then, we only consider misspellings that cannot be justified as abbreviations. Finally, we compute for every region the proportion $\varepsilon_i$ of {\it misspellers} among the Twitter population. If the fraction of misspellers per geographical area is a proxy for the educational level of that region, we expect a positive correlation between $\varepsilon_i$ and unemployment. Indeed we find (see figure \ref{fig:corr_unemployment}) that there is a strong correlation between the fraction of misspellers and unemployment.
	\item {\it Social media interactions and geographical flow diversity}: following the ideas in \cite{Eagle21052010} which correlated the economical development of an area with the diversity of communications with other areas, we consider all tweets mentioning another user and take them as a proxy for communication between users. Then we compute the number of communications $w_{ij}$ between areas $i$ and $j$ as the number of mentions between users in those areas. To measure the diversity we use as in \cite{Eagle21052010} the informational normalized entropy (Entropy 1) $S_{u,i} = -\sum_j p_{ij} \log p_{ij} / S_{r,i} $ where $p_{ij} = w_{ij} / \sum_{j} w_{ij}$ and (Entropy 2) $S_{r,i} = \log k_i$ with $k_i$ the number of different areas with which users in area $i$ have interacted. As in \cite{Eagle21052010}, we find that areas with large unemployment have less diverse communication patterns than areas with low unemployment. This translates in a strong negative correlation between $S_i$ and the unemployment, see figure \ref{fig:corr_unemployment}. Similar ideas are applied to the flows of people between areas to investigate the diversity of the geographical flows through the entropy $\tilde S_i = -\sum_j \tilde p_{ij} \log \tilde p_{ij} / \tilde S_{r,i}$, where $\tilde p_{ij} = T_{ij} / \sum_j T_{ij}$ and $S_{r,i} = \log(\tilde k_i)$ with $\tilde k_i$ the number of different areas which has been visited by users that live in area $i$. Figure \ref{fig:corr_unemployment} shows that as in \cite{smith2013finger}, correlation of these geographical entropies is low with economical development.
\end{itemize}
Normalization of variables is discussed in SI section 5. We have also studied the correlation between the variables considered. As expected, variables in each group show moderate correlations between them. However the inspection of the correlation matrix and a Principal Component Analysis of the variables considered, show that there is information (as percentage of variance in the data) in each of the groups of variables, see SI section 5. Because of these two facts we restrict our analysis to the variables within each group with the highest correlation with the unemployment, namely the penetration rate $\tau_i$, the social and mobility diversity variables $S_{u,i}$ and $\tilde S_{u,i}$, the morning activity $\nu_{\mathrm{mrng},i}$, the fraction of misspellers $\varepsilon_i$ and fraction of {\em employment}-related tweets $\mu_{emp,i}$. 


\section{Explanatory power of social media in unemployment}
The four previous groups of variables are fingerprints of human behavior reflected on the Twitter usage habits. As we observed in figure \ref{fig:corr_unemployment}, all of them exhibit statistically strong correlations with unemployment. The question we address in this section is whether those variables suffice to explain the observed unemployment (their explanatory power) and also determine the most important ones among themselves (which give more explanatory power than others). Note that we are not stating a causality arrow between the measures built in the previous section and the unemployment rate but only exploring whether they can be used as alternative indicators with a real translation in the economy. 

\begin{figure*}[!ht]
\centering
	\includegraphics[width=\textwidth]{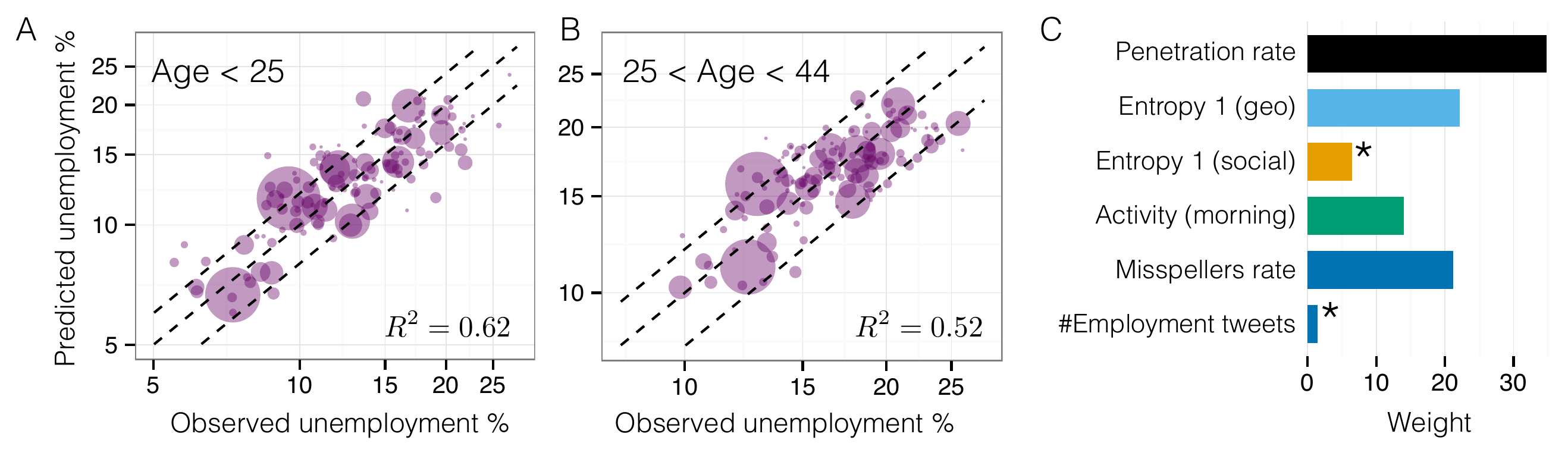}
	\caption{A) and B) Performance of the model, showing the predicted unemployment rate for ages below 25 versus the observed one, $R^2=0.62$ and with ages between 25 and 44. Dashed lines correspond to the equality line and $\pm 20\%$ error. C) Percentage of weight for each of the variables in the regression model using the relative weight of the absolute values of coefficients in the regression model (see SI section 10). Variables marked with $*$ are not statistical significant in the model.}
	\label{fig:model_unemployment}
\end{figure*}

Figure \ref{fig:model_unemployment} shows the result of a simple linear regression model for the observed unemployment for ages below 25 years as a function of the variables which have more correlation with the unemployment. The model has a significant $R^2 = 0.62$ showing that there is a large explanatory power of the unemployment encoded in the behavioral variables extracted from Twitter. However, not all the variables weight equally in the model: specifically, the penetration rate, geographical diversity, morning activity and fraction of misspellers account for up to 92\% of the explained variance, while social diversity and number of {\em employment} related tweets are not statistical significant (see SI section 10 for the methods used to determine the relative importance of the variables). It is interesting to note that while social diversity obtained by mobile phone communications was a key variable in the explanation of deprivation indexes in \cite{Eagle21052010,smith2013finger}, the communication diversity of twitter users seem to have a minor role in the explanation of heterogeneity of unemployment in Spain. 

Similar explanatory power is found for other age groups: $R^2 = 0.44$ for all ages and $R^2 = 0.52$ for ages between 25 and 44 years. However, the model degrades for ages above 44 years ($R^2 = 0.26)$ proving that our variables mainly described the behavior of the most represented age groups in Twitter, namely those below 44 years old. On the other hand, since our Twitter variables seem to describe the behavior of young people, we have investigated whether Twitter constructed variables have similar explanatory value (in terms of $R^2$) than simple census demographic variables for young people. However, regression models including young population rate yield to a minor improvement $R^2 = 0.65$, while young population rate only gives $R^2=0.24$, a result which shows that Twitter variables do indeed posses a genuine explanatory power away from their simple demographic representation. Finally, our model have the largest explanatory power for detected communities, but large $R^2$ are also found for other geographical areas like counties ($R^2 = 0.54$) and provinces ($R^2=0.65$), see SI section 9.

\section{Discussion}
This work serves as a proof of concept for how a wide range of behavioral features linked to socioeconomic behavior can be inferred from the digital traces that are left by publicly-available social media. In particular, we demonstrate that behavioral features related to unemployment can be recovered from the digital exhaust left by the microblogging network Twitter. First of all, Twitter geolocalized traces, together with off-the-shelve community detection algorithms, render an optimal partition of a country for economical activity, showing the remarkable power of social media to understand and unveil economical behavior at a country-scale. This insight is likely to apply to other administrative definitions in other countries, specially when considering large cities with an inherent dynamical nature and evolution of mobility fluxes, and cities composed of small satellite cities with arbitrary agglomerations or division among them (e.g. London, NYC, Singapore). This result is unsurprising: it should be natural to recompute city clusters/communities of activity based on their real time mobility, which may vary considerably faster than the update rates of mobility and travel surveys~\cite{Barthelemy20111,expert2011uncovering,sobolevsky2013delineating}. 

Our main result demonstrates that several key indicators, different penetration rates among regions, fingerprints of the temporal patterns of activity, content lexical correctness and geo-social connectivities among regions, can be extracted from social media, and then used to infer unemployment levels. These findings shed light in two directions: first, on how individuals' extensive use of their social channels allow us to characterize cities based on their activity in a meaningful fashion and, secondly, on how this information can be used to build economic indicators that are directly related to the economy. Regarding the latter, our work is important for understanding how country-scale analysis of Social Media should consider the demographic but also the economical difference between users. As we have shown, users in areas with large unemployment have different mobility, different social interactions, and different daily activity than those in low unemployment areas. This intertwined relationship between user behavior and employment should be considered not only in economical analysis derived from social media, but also in other applications like marketing, communication, social mobilization, etc.

It is particularly remarkable that Twitter data can provide these accurate results. Twitter is, among the many currently popular social networking platforms, perhaps the noisiest, sparsest, more `sabotaged' medium: very few users send out messages at a regular rate, most of the users do not have geolocated information, the social relationships (followers/followers) contains a lot of unused/unimportant links, it is plagued by spam-bots, and last but not least, we have no way to identify the motive/goal/functionality of mobility fluxes we are able to extract. These limitations are not particular to our sample, but general to the sample Twitter data being employed in the computational social science community. Despite all these caveats, we are able to show that even some simple filtering techniques together with basic statistical regressions yields predictive power about a variable as important as unemployment. Other social media platforms such as Facebook, Google+, Sina Weibo, Instagram, Orkut, or Flicker with more granular and consistent individual data are likely to provide similar or better results by themselves, or in combination. Further improvements can be obtained by the use of more sophisticated statistical machine learning techniques, some of them even tailored to the peculiarities of social media data. Our work serves to illustrate the tremendous potential of these new digital datasets to improve the understanding of society's functioning at the finer scales of granularity.

The usefulness of our approach must be considered against the cost and update rate of performing detailed surveys of mobility, social structure, and economic performance. Our database is publicly articulated, which means that our analysis could be replicated easily in other countries, other time periods and with different scopes. Naturally, survey results provide more accurate results, but they also consume considerably higher financial and human resources, employing hundreds of people and taking months, even years to complete and be released --- they are so costly that countries going through economic recession have considered discontinuing them, or altering their update rate in recent times. A particularly problematic aspect of these surveys is that they are ``out-of-sync'' i.e. census may be up to date, whereas those same individuals' travel surveys may not be, and therefore drawing inferences between both may be particularly difficult. This is a particularly challenging problem that the immediateness of social media can help ameliorate. 

A few questions remain open for further investigation. How can traditional surveys and social media digital traces be best combined to maximize their predictive ability? Can social media provide a reliable leading indicator to unemployment, and in general, economic surveys? How much reliable lead is it possible, if at all? As we have found, Twitter penetration and educational levels are found to be correlated with unemployment, but this levels are unlikely to change rapidly to describe or anticipate changes in the economy or unemployment. However, other indicators like daily activity, social interactions and geographical mobility are more connected with our daily activity and perhaps they have more predicting power to show and/or anticipate sudden changes in employment. The relationship between unemployment and individual and group behavior may help contextualize the multiple factors affecting the socioeconomic well-being of a region: while penetration, content, daily activity and mobility diversity seem to be highly correlated to unemployment in Spain, different weights for each group of traces might be expected in other countries \cite{Eagle21052010}. Finally, digital traces could serve as an alternative (some times the only one available) to the lack of surveys in poor or remote areas \cite{smith2013ubiquitous,Soto:2011:PSL:2021855.2021893}. Another interesting avenue of research involves the use of social media to detect mismatches between the real (hidden, underground) economy and the officially reported~\cite{schneider2011shadow}. 

Most importantly, the immediacy of social media may also allow governments to better measure and understand the effect of policies, social changes, natural or man-made disasters in the economical status of cities in almost real-time \cite{lazer2009life,rutherford2013limits}. These new avenues for research provide great opportunities at the intersection of the economic, social, and computational sciences that originate from these new widespread inexpensive datasets.

\section*{Acknowledgments}
We would like to thank Kristina Lerman, Lada Adamic, James Fowler, Daniel Villatoro, and Ricardo Herranz for stimulating discussions, and Yuri Kryvasheyeu and Thomas Bochynek for their critical reading of the manuscript.
This work was partially supported by 
Spanish Ministry of Science and Technology Grant FIS2013-47532-C3-3-P (to A. L., M. G. H. and E. M.). Manuel Cebrian is funded by the Australian Government as represented by the Department of Broadband, Communications and Digital Economy and the Australian Research Council through the ICT Centre of Excellence program.

\bibliographystyle{pnas2009}
\small
\bibliography{movilidadbib}

\newpage

\appendix

\ \ \ 
\newpage

\twocolumn[\begin{@twocolumnfalse}

\noindent {\bf \Large Supporting Information for}\\

\noindent{\LARGE \em Social media fingerprints of unemployment}

\vspace{0.1cm}

\noindent
{Alejandro Llorente, Manuel Garc\'{\i}a-Herr\'anz Manuel Cebri\'an, and Esteban Moro}

\vspace{0.2cm}
\hrule
\vspace{0.4cm}

\end{@twocolumnfalse}]

\singlespace
\section*{S1. The dataset}\label{SIs1}
Twitter provides an extremely rich and publicly available data set of user interactions, information flows and, thanks to the geo location of tweets, user movements. Nevertheless, the representativeness of this geo-located Twitter as a global source of mobility data has still received sparse attention. In this sense, while \cite{SIhawelka2013geo}  present a promising and extensive study regarding global country-to-country movements (mostly driven by tourism), within-country human flows (comprising not only internal tourism but also,  in a greater extent than country-to-country travels, visiting and commuting) still need further investigation. Therefore, throughout this work we will compare our findings using geo-located Twitter with similar study using commuting surveys.
         
For the Twitter analysis, we consider almost 146 million geo-located 
Twitter messages (tweet(s)), collected through the public API provided by Twitter for the
continental part of Spain and from 29th November 2012 to 10th April
2013. In this dataset we consider that there has been a trip from
place $l$ to place $k$ if a user has tweeted in place $l$ and place
$k$ consecutively. We only keep those transitions when the first tweet
and the second one are dated in the same day. We filter the trips
database to avoid unrealistic transitions and keep only trips with
a geographical displacement larger than 1km (See Methods section). By
this method, 1.38 million of trips from 167,376 different users are
considered in our work. 

From those trips we construct the mobility flow $T_{ij}$
between municipalities, which measures the number of trips in our
database in which the origin is within city $i$ boundaries and destination lies
within  those of city $j$.

We also consider population and economical information about the municipalities from
the Spanish Census (2011) \cite{SIcenso} and unemployment figures from the Public Service of Employment (Servicio P\'{u}blico de Empleo Estatal, SEPE) \cite{SIsepe}. In the former In the latter case, registered unemployment (in number of persons) is given for each Spanish municipality by gender, age, and month. To get unemployment rates we divide register unemployment by the total workforce in the municipality, estimated as the number of people with age between 16 and 65 years.

\section*{S2. Twitter as mobility proxy}

\begin{figure*}[t]
	\includegraphics[width=1.01\textwidth]{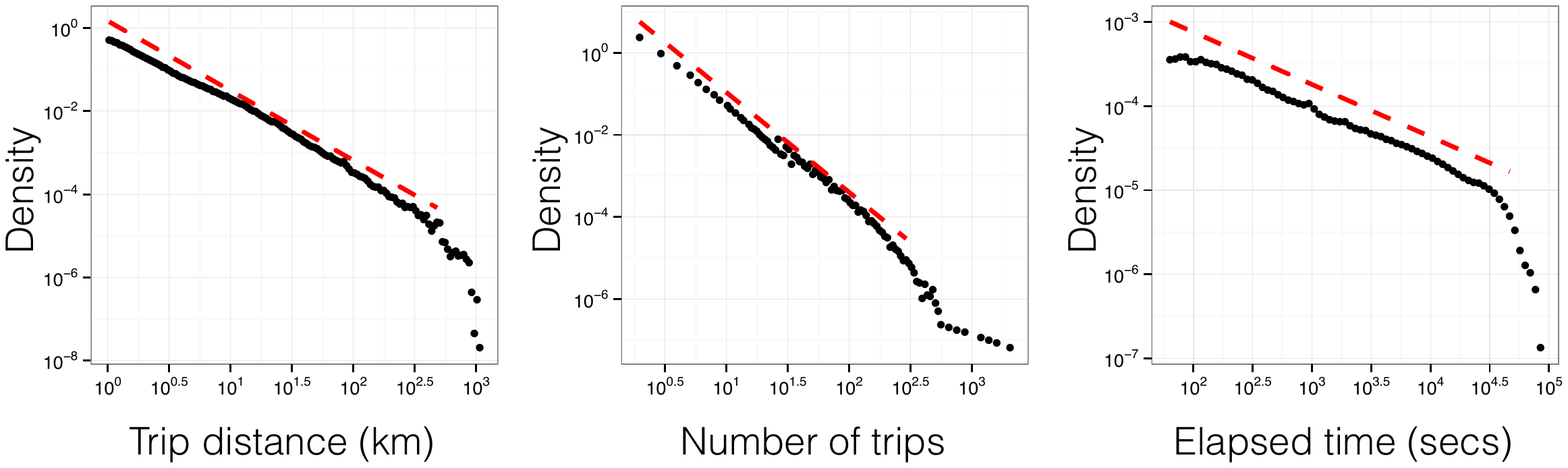}
	\caption{Probability distributions for the different properties of daily trips in the Twitter dataset. Dashed lines corresponds to a power law fit with exponents $-1.67$, $-2.43$ and $-0.62$ respectively}
	\label{fig:trips_stats}
\end{figure*}

Considering all of the available transitions in our database, one can compute the distance between origin and destination, the elapsed time of the transition and the number of trips per user among many other statistics. All of them seems to show a Power-law distribution with a cutoff due to the finite spatial size of Spain and the constraint of considering only transitions where the origin and destination checkins are done the same day. Focusing on the log-linear part of the distributions, self-similar behaviors arise when Twitter based mobility is analyzed (see figure \ref{fig:trips_stats}). 

Twitter based inter-city flows can be well modelled by means of the The Gravity Law, which is one of the most extended methods to represent human mobility \cite{SIashtakala1987generalized,SIschneider1959gravity} , with applications in many fields like urban planning \cite{SIwilson1974urban}, traffic engineering \cite{SIcasey1955law} or transportation problems \cite{SIevans1973relationship}. Gravity Law is also the solution to the problem of maximizing the entropy of the particle distribution among all the possible trips using statistical mechanics techniques \cite{SIwilson1970entropy,SIbierlaire1997mathematical}. Recently, it has also been used as a model for human mobility based on cell phone traces \cite{SIsimini2012universal, SIexpert2011uncovering, SIsong2010modelling} and social media data at a global scale \cite{SIhawelka2013geo} and at the inter-city level \cite{SIliu2013uncovering}.

The Gravity Model for human mobility assume that the flows between 
cities can be explained by the expression 
\begin{equation}\label{eq:SIgravity}
	T_{ij}^{grav}=\frac{P_i^{\alpha_1} P_j^{\alpha_2}}{d_{ij}^\beta}
\end{equation}
where $T_{ij}^{grav}$ is the flow, in terms of number of people, 
between cities $i$ and $j$, $d_{ij}$ is the geographical distance and
$P_i$ and $P_j$ the population of every city respectively. 
	
Given the data we can obtain the parameters of the model by Weighted
Least Squares Minimization,
\begin{equation}
	\alpha_{1}^{*},\alpha_{2}^{*},\beta^{*} = \underset{\alpha_{1},\alpha_{2},\beta}{\operatorname{argmin}}
	\frac{1}{N} \sum_{i,j} w_{ij} \left( T_{ij} - T_{ij}^{grav} \right)^2
\end{equation}
where $N$ is the total number of connections in the mobility graph and 
$w_{ij}$ is a weight proportional to the number of observed transitions between
$i$ and $j$. In particular we find that taking $w_{ij} = T_{ij}^{1.3}$ gives the best performance in the model.

In our case, this model fits quite accurately the inter-city mobility based on Twitter GPS checkins (see table \ref{tab:grav_model_params}). Even though we are considering $T_{ij}$ not necessarily symmetric, the exponents of the populations are similar indicating that we are observing a similar flows in both directions between $i$ and $j$.

\begin{table*}[h!]\centering
	\begin{tabular}{ |c|c|c| }
		\hline
			\multicolumn{3}{|c|}{{\bf Gravity Model}} \\
  		\hline
		\hline
		Parameter & Description & Spain \\ 
		\hline
		$\alpha_1$ & Origin exponent & $0.477^{***} (0.002)$ \\ 
		$\alpha_2$ & Destination exponent & $0.478^{***} (0.002)$\\ 
		$\beta$ & Distance exponent & $1.05^{***} (0.0035)$  \\ 
		\hline
		\hline
  		$R^2$ & Goodness of fit & 0.797 \\ 
		$\phi$ & Correlation between $T_{ij}$ and $T_{ij}^{gra}$ & 0.826 \\ 
		\hline
	\end{tabular}
	\caption{Description of the parameters for the Gravity Law Model in geo-tagged 
	social media data for Spain. ($***$) means significance $p < 0.0001$. }
	\label{tab:grav_model_params}
\end{table*}

\section*{S3. Community structures in inter-city mobility graph}\label{sec:community}

\begin{figure*}[t]
	\includegraphics[width=\textwidth]{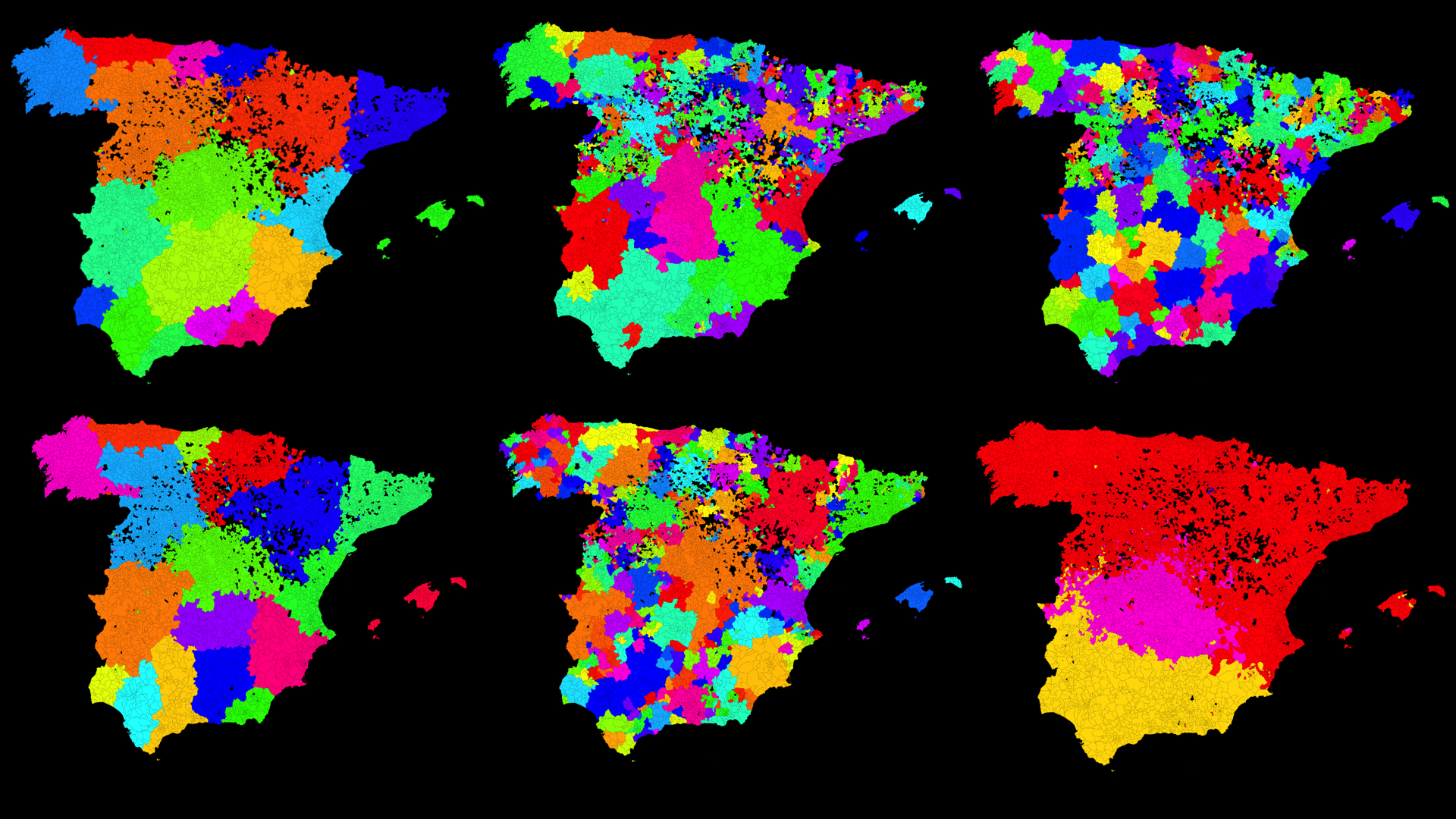}
	\caption{From left to right and from top to bottom: Fastgreedy, Walktrap, Infomap, Multilevel, Label Propagation and Leading Eigenvector communities on Twitter based mobility transitions.}
	\label{fig:comms_plot}
\end{figure*}

Typically, complex networks exhibit community structure, that is, there are subsets of nodes that are more densely connected among them comparing to the rest of the nodes. In mobility networks, whose nodes correspond to geographical areas, these communities are interpreted as zones with high common activity and tend to be constrained by geographical and political barriers. We check whether this is also observed in our dataset by performing 6 state-of-art community detection algorithms: FastGreedy \cite{SIclauset2004finding}, Walktrap \cite{SIpons2005computing}, Infomap \cite{SIrosvall2008maps}, MultiLevel \cite{SIblondel2008fast}, Label Propagation \cite{SIraghavan2007near} and Leading Eigenvector \cite{SInewman2006finding}. These six different algorithms exhibit different community structures in terms of number of communities, average size of community or modularity (see table \ref{tab:comms_provs_counties}). Members (municipalities) of the resulting communities are spatially connected except some few cases as figure \ref{fig:comms_plot} shows. We test the statistical robustness of the obtained communities by randomly removing a proportion $p$ of the original links and performing the algorithms on this new graph $G_{p}$. We will consider that communities are robust when the communities given for the original network $G$ and $G_p$ are highly similar. In order to compare two arbitrary memberships to communities, we use the Normalized Mutual Information (NMI) method described in \cite{SIdanon2005comparing} which returns 0 when two memberships are totally different and 1 when we compare two equal memberships. We compute the NMI for each chosen algorithm performed on $G$ and $G_{p}$, for $p$ between $1\%$ and $10\%$, concluding that obtained community structures are robust because they are not broken when some randomly chosen links are removed (see table \ref{tab:nmi_gp_g}).

\begin{figure}[h!]
 \centering
	\includegraphics[width=0.5\textwidth]{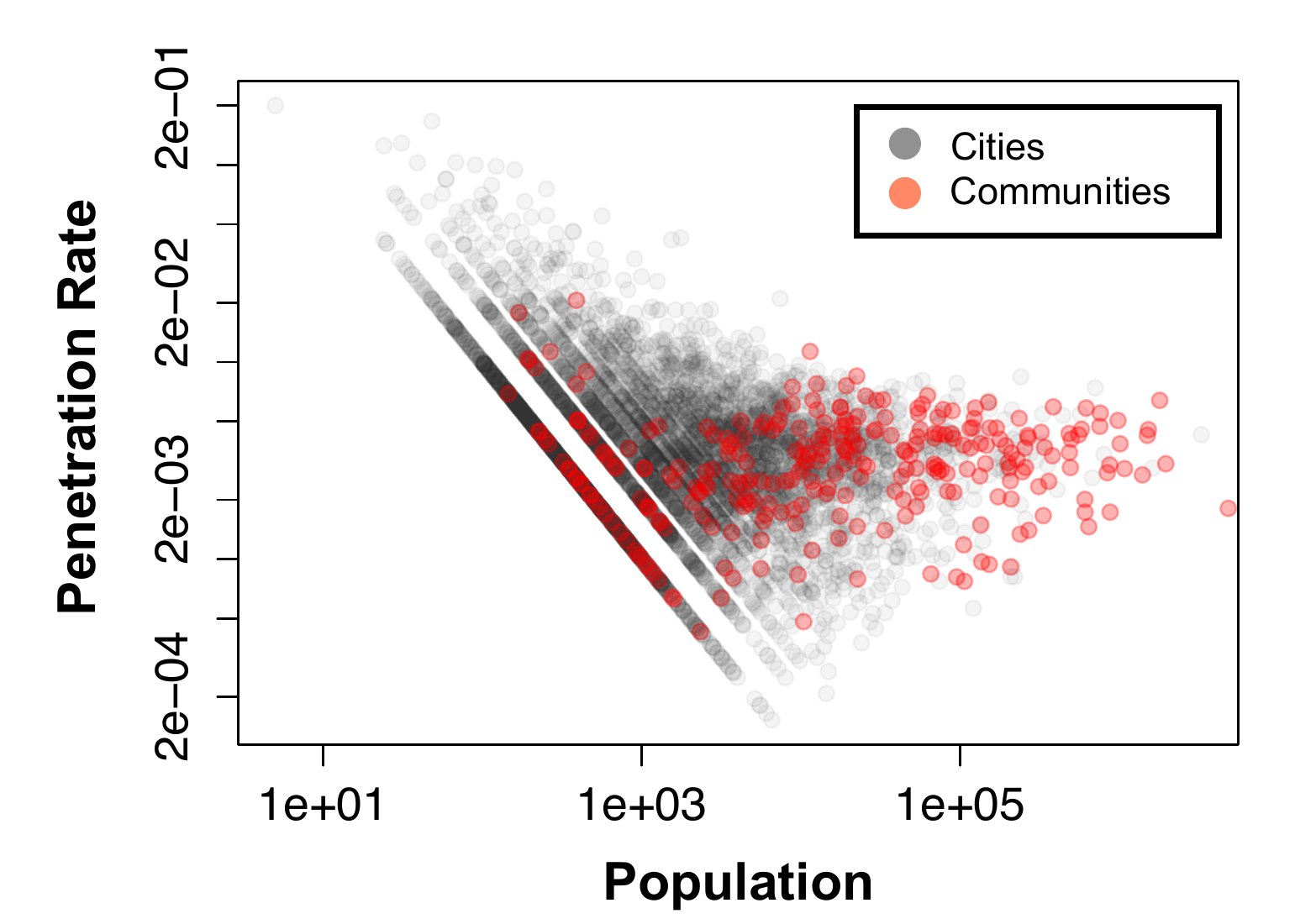}
	\caption{Penetration rates for both cities and detected communities.}
	\label{fig:comms_twpen}
\end{figure}

\begin{table*}
	\centering
	\begin{tabular}{ |c|c|c|c|c|c|c|c|c|c|c| }
		\hline
			\multicolumn{11}{|c|}{{\bf NMI between $G$ and $G_p$ for different $p$}} \\
  		\hline
		\hline
Algorithm& $p=0.01$ & $0.02$ & $0.03$ & $0.04$ & $0.05$ & $0.06$ & $0.07$ & $0.08$ & $0.09$ & $0.1$\\ 
\hline 
FG & $0.995$ & $0.992$ & $0.989$ & $0.983$ & $0.981$ & $0.977$ & $0.983$ & $0.969$ & $0.980$ & $0.959$ \\ 
WT & $0.954$ & $0.959$ & $0.950$ & $0.954$ & $0.945$ & $0.948$ & $0.947$ & $0.935$ & $0.926$ & $0.931$ \\  
IM & $0.988$ & $0.981$ & $0.980$ & $0.981$ & $0.978$ & $0.974$ & $0.975$ & $0.970$ & $0.969$ & $0.966$ \\ 
ML & $0.994$ & $0.978$ & $0.979$ & $0.983$ & $0.948$ & $0.934$ & $0.972$ & $0.952$ & $0.973$ & $0.947$ \\  
LP & $0.906$ & $0.908$ & $0.911$ & $0.915$ & $0.895$ & $0.907$ & $0.907$ & $0.893$ & $0.905$ & $0.904$ \\  
LE & $0.960$ & $0.957$ & $0.956$ & $0.859$ & $0.910$ & $0.892$ & $0.908$ & $0.858$ & $0.885$ & $0.884$ \\ 
		\hline
	\end{tabular}
	\caption{NMI measure comparing $G$ and $G_p$.}
	\label{tab:nmi_gp_g}
\end{table*}

As other works have shown, mobility graph communities are usually interpreted in terms of geographical and political barriers and a natural question is whether the mobility based communities are related to any of these barriers. In Spain, there are different territorial divisions for administration purposes. In this work, we consider two of them: provinces, defined in 1978 Constitution, are 48 different heterogeneous aggregations of municipalities; and counties ({\it comarca} in Spanish terminology) which are traditional aggregations of municipalities mainly based on Spanish holography (rivers, valleys, ridges, etc) and some of them are composed by municipalities of different provinces. We use again the NMI method to compare the communities structure given by the algorithms to the administrative limits. Except Leading Eigenvector algorithm, the rest of methods return communities that are quite related to provinces ($NMI \approx 0.7$) whereas for the county administration limits, higher variability is observed. In this last case, the algorithm providing more relationship with county limits is Infomap, $NMI \approx 0.83$. Therefore, Twitter based mobility summarizes the inter-city flows exhibiting that these flows are influenced by geographical and political barriers.

\begin{table*}[h!]
	\centering
	\begin{tabular}{|>{\centering\arraybackslash}p{1.5cm}|>{\centering\arraybackslash}p{2cm}|>{\centering\arraybackslash}p{2cm}|>{\centering\arraybackslash}p{1.7cm}|>{\centering\arraybackslash}p{1.7cm}|>{\centering\arraybackslash}p{1.7cm}|>{\centering\arraybackslash}p{1.7cm}|}
	\hline
	\multicolumn{7}{|c|}{~~{\bf Communities Stats}~~} \\
	\hline
	Algorithm & $\langle|N_i|\rangle_i$ & $\max \{ |N_i| \}$ & $|\{N_i\}|$ & Modularity & {\it NMI P} & {\it NMI C} \\
	\hline
	\hline
	FG & 309.696 & 1385 & 23 & 0.726 & 0.712 & 0.590 \\
	WT & 9.262 & 433 & 769 & 0.417 & 0.744 & 0.757 \\
	IM & 21.011 & 143 & 339 & 0.758 & 0.770 & 0.831 \\
	ML & 323.772 & 1132 & 22 & 0.800 & 0.717 & 0.599 \\
	LP & 22.052 & 750 & 323 & 0.732 & 0.749 & 0.761 \\
	LE & 1017.571 & 5344 & 7 & 0.381 & 0.264 & 0.205 \\
	\hline
	\end{tabular}
	\caption{Statistics of the communities $\{N_i\}$ returned by the six algorithms. {\it NMI P} refers to the comparison between communities and provinces whereas {\it NMI C} considers counties instead of provinces.}
	\label{tab:comms_provs_counties}
\end{table*}

\section*{S4. Twitter demographics and unemployment rates}

\begin{figure}[th]
	\centering
	\includegraphics[width=0.45\textwidth]{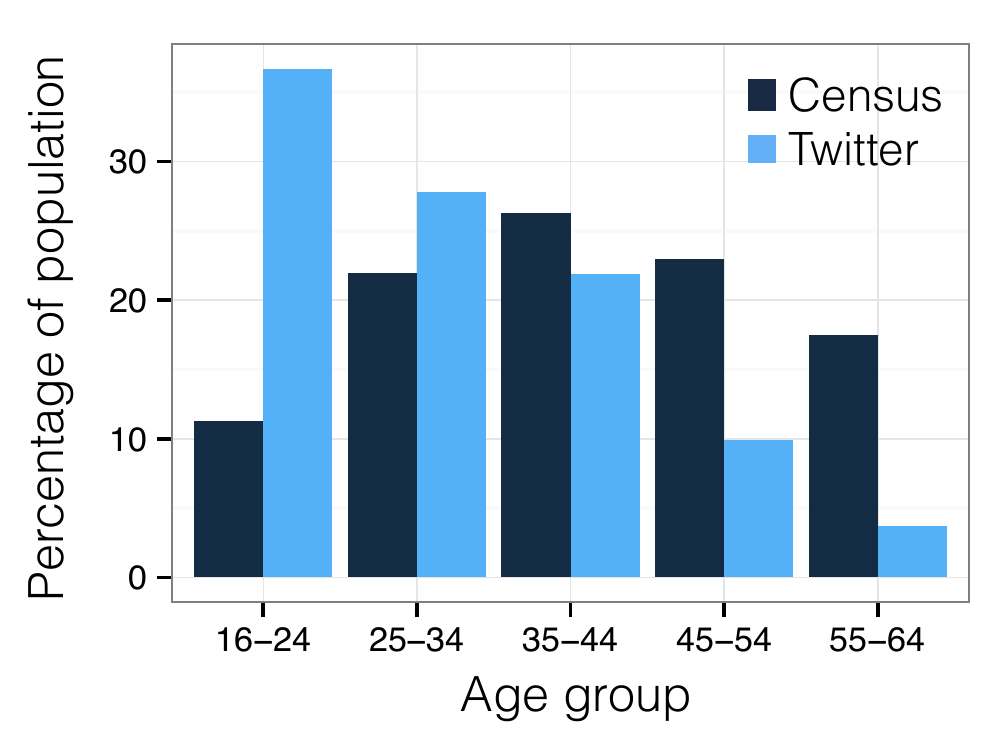}\ \ \ \ \ \ \ 
	\includegraphics[width=0.45\textwidth]{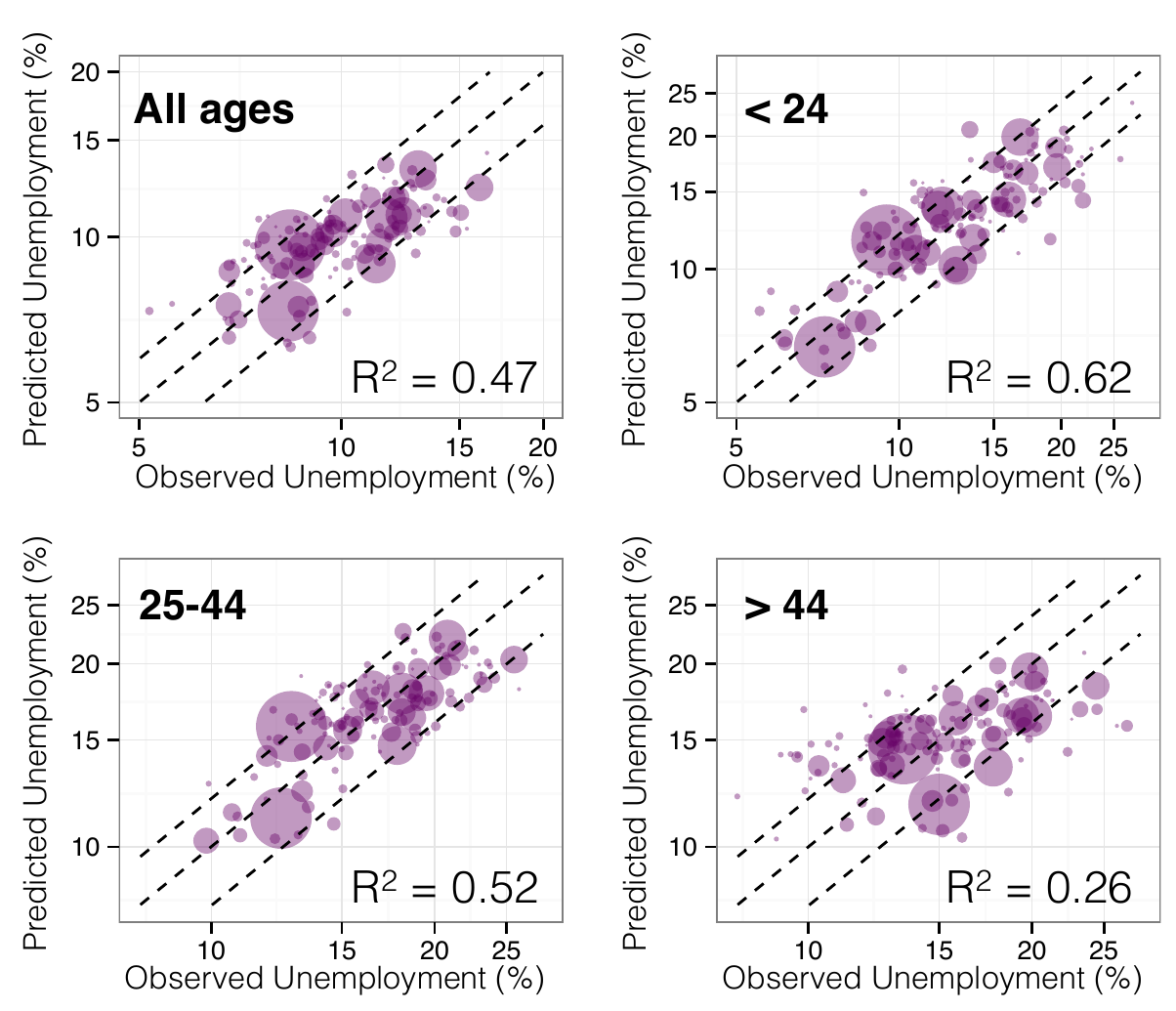}
	\caption{Top: Percentage of population in each age group from the Spanish Census (dark bars) and surveys about users in Twitter (light bars). Bottom: performance of the linear models for each of the age groups.}
	\label{fig:census}
\end{figure}

Different age groups are not equally represented in Twitter. Recent surveys (2012) in Spain suggest that most (86\%) of users in Twitter are 16 to 44 years old. Comparison of the percentage of users per age group with the total population within the same groups (see figure \ref{fig:census}) reveals that groups of ages above 35 years old are under-represented in Twitter. Thus our Twitter data will be more revealing when trying to describe unemployment in age groups below 44 years old. This is indeed what we find when we try to build a linear model for the rate unemployment in different age groups with the same Twitter variables: while unemployment rates for ages below 24 can be fitted to a linear model with $R^2 = 0.62$ we find that regression models for unemployment rates for ages between 25 and 44 have a $R^2 = 0.52$, while for ages above 44 we get only $R^2 = 0.26$. Table \ref{table:agegroups} summarizes the results for the regression models of unemployment rates in each age group, showing that our Twitter variables have more explanatory power for ages below 44. Finally, in figure \ref{fig:census} we can see the performance of the model at different age groups and, once again, it is obvious the poor explanatory power of the Twitter variables for the unemployment rate in ages above 44 years old. 

\begin{table*}[!h]
\begin{center}
\begin{tabular}{l c c c c c}
\hline
                  	& All ages		& $< 24$ 		& $25-44$ 		& $>44$ \\
\hline
(Intercept)       	& $0.11*{***}$	& $0.10^{***}$	& $0.20^{***}$ 	& $0.20^{***}$	\\
 				  	& $(0.02)$		& $(0.03)$ 		& $(0.03)$ 		& $(0.035)$		\\
Penetration rate    & $3.23^{*}$	& $8.57^{***}$	& $6.28^{**}$ 	&$2.40$		\\
					& $(1.41)$		& $(2.22)$ 		& $(2.17)$		&$(2.77)$		\\
Geographical diversity &$0.03$		& $0.15^{***}$	& $0.08^{*}$ 	&$0.06$		\\
					& $(0.02)$		& $(0.04)$		& $(0.04)$		&$(0.05)$		\\	
Social diversity 	& $-0.03^{*}$	& $-0.03$  		& $-0.05^{*}$  	&$-0.06^{*}$      	\\
					& $(0.01)$		& $(0.02)$		& $(0.02)$		&$(0.03)$\\
Morning activity    & $-0.69^{*}$	& $-1.30^{**}$	& $-1.53^{***}$	&$-1.19^{*}$	 	\\
					& $(0.26)$		& $(0.42)$		& $(0.41)$		&$(0.52)$		\\
Misspellers rate    & $11.56$		& $31.51^{*}$  	& $15.46$  		&$23.60$		\\
					& $(8.13)$		& $(12.78)$		& $(12.48)$		&$(15.94)$		\\
\emph{Employment} mentions  & $-1.80$ & $3.17$		& $-9.94$		&$2.71$  	\\
					& $(6.27)$		& $(9.86)$		& $(9.64)$		&$(12.3)$		\\ 
\hline
R$^2$             	& $0.47$		&$0.64$		& $0.55$		& $0.29$		\\
Adj. R$^2$        	& $0.44$		&$0.62$		& $0.52$ 		& $0.26$		\\
\hline
\multicolumn{4}{l}{\small{$^{***}p<0.001$, $^{**}p<0.01$, $^*p<0.05$}}
\end{tabular}
\caption{Regression table for the different models in which unemployment for different age groups is fitted. The {\em All ages} model is the fit to the general rate of unemployment in each geographical area, while the other models are for the rates of unemployment in groups of less than 24 years, between 25 and 44 years and above 44 years.}
\label{table:agegroups}
\end{center}
\end{table*}

\section*{S5. Properties of Twitter variables}

\subsection*{Normalization and distributions}
\begin{figure*}[ht!]
	\centering
	\includegraphics[width=0.85\textwidth]{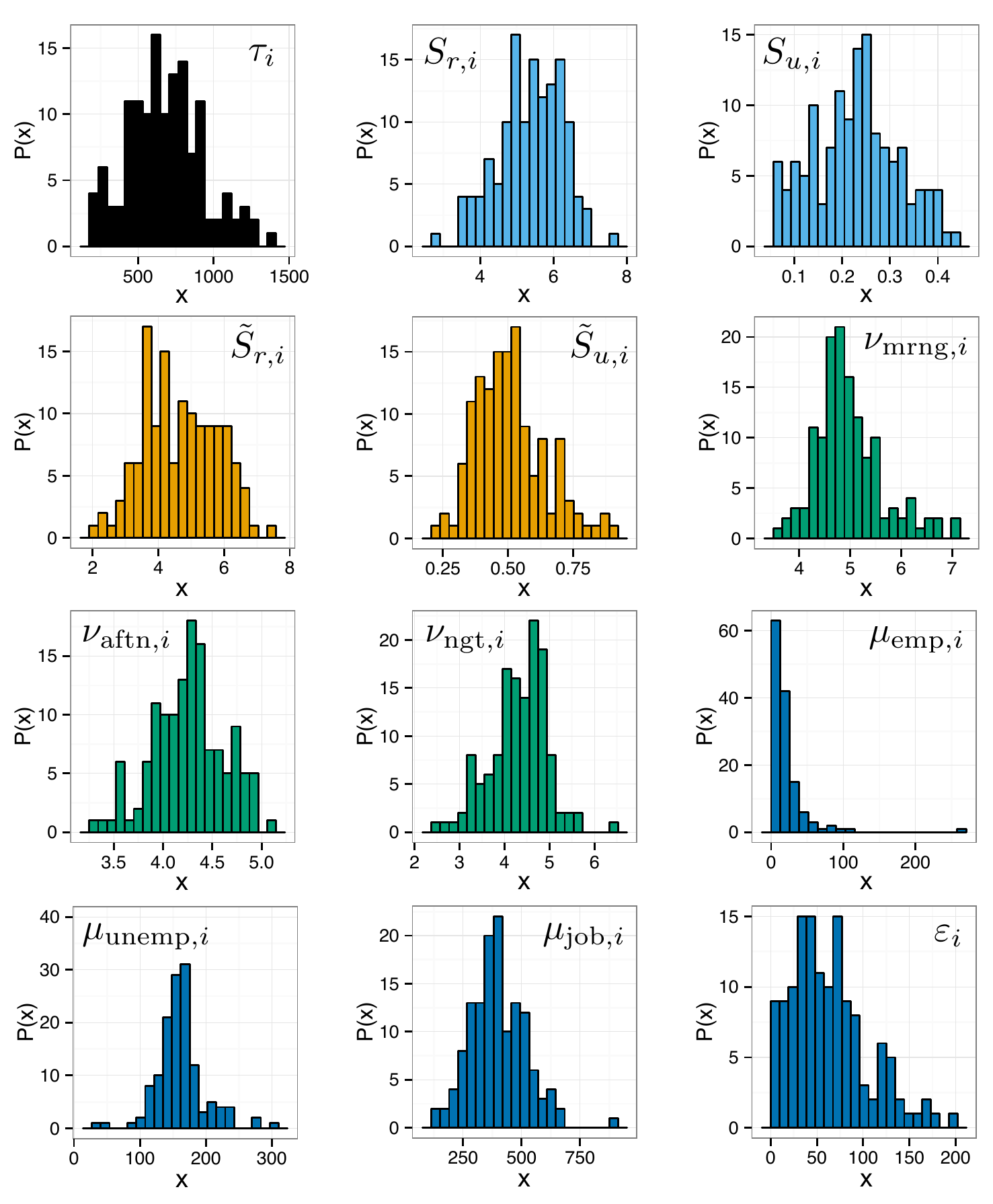}
	\caption{Frequency plots for each variable constructed from Twitter.}
	\label{fig:histograms}
\end{figure*}

Heterogeneity between the values of variables constructed from Twitter is large but moderate, as histograms in figure \ref{fig:histograms} show. We did not find any geographical area with anomalous values in any of the variables considered. Variables are normalized in different ways: both the penetration $\tau_i$ and misspellers rate $\varepsilon_i$ are defined as the number of users or misspellers per 100.000 persons (population); activity variables $\nu_i$ are normalized as the percentage of tweets per time interval; finally, number of tweets that mention a specific term $\mu_{i}$ are also given per 100.000 tweets published in the geographical area.

\subsection*{Correlation between variables}
\begin{figure}[!h]
	\centering
	\includegraphics[width=0.45\textwidth]{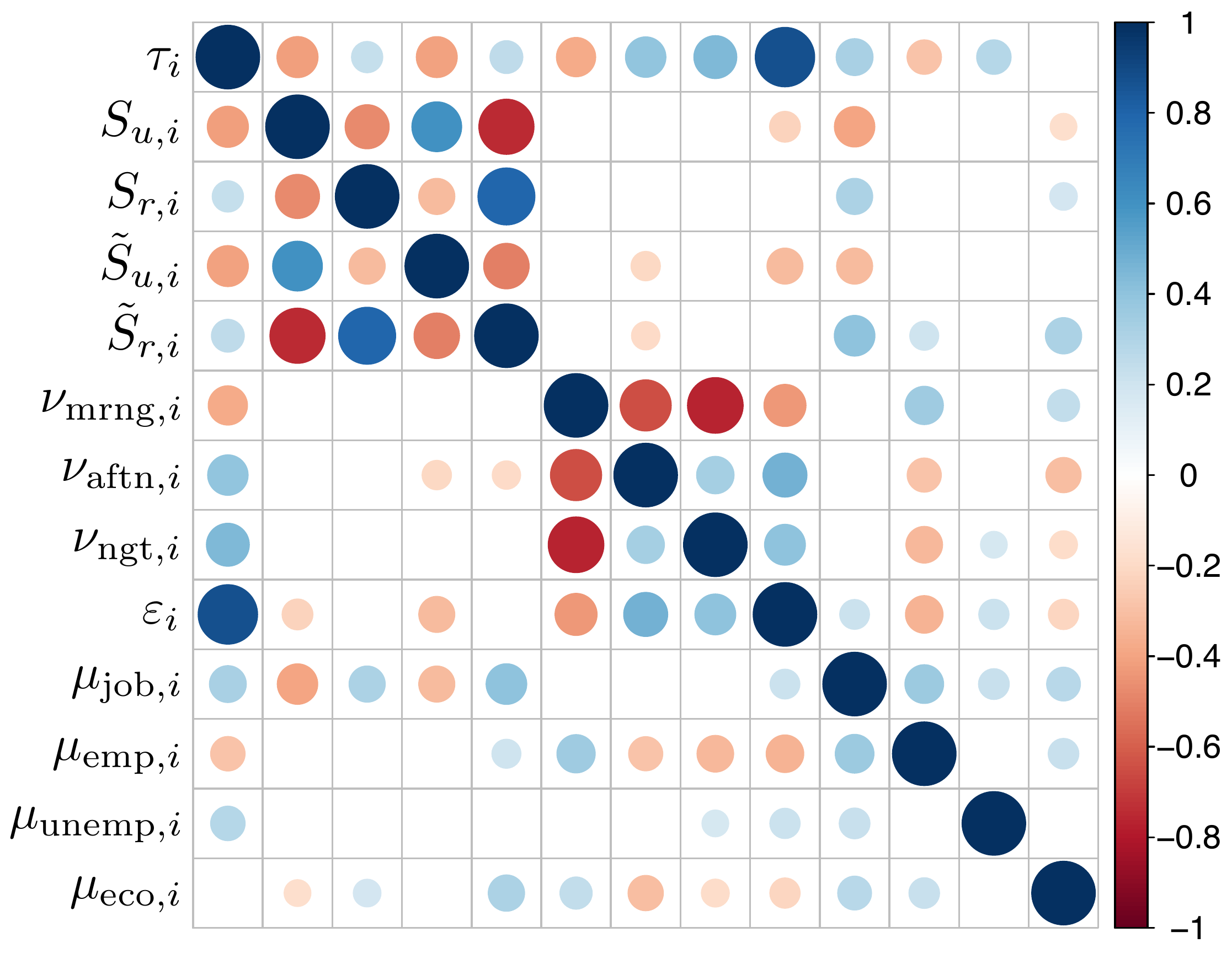}
	\includegraphics[width=0.43\textwidth]{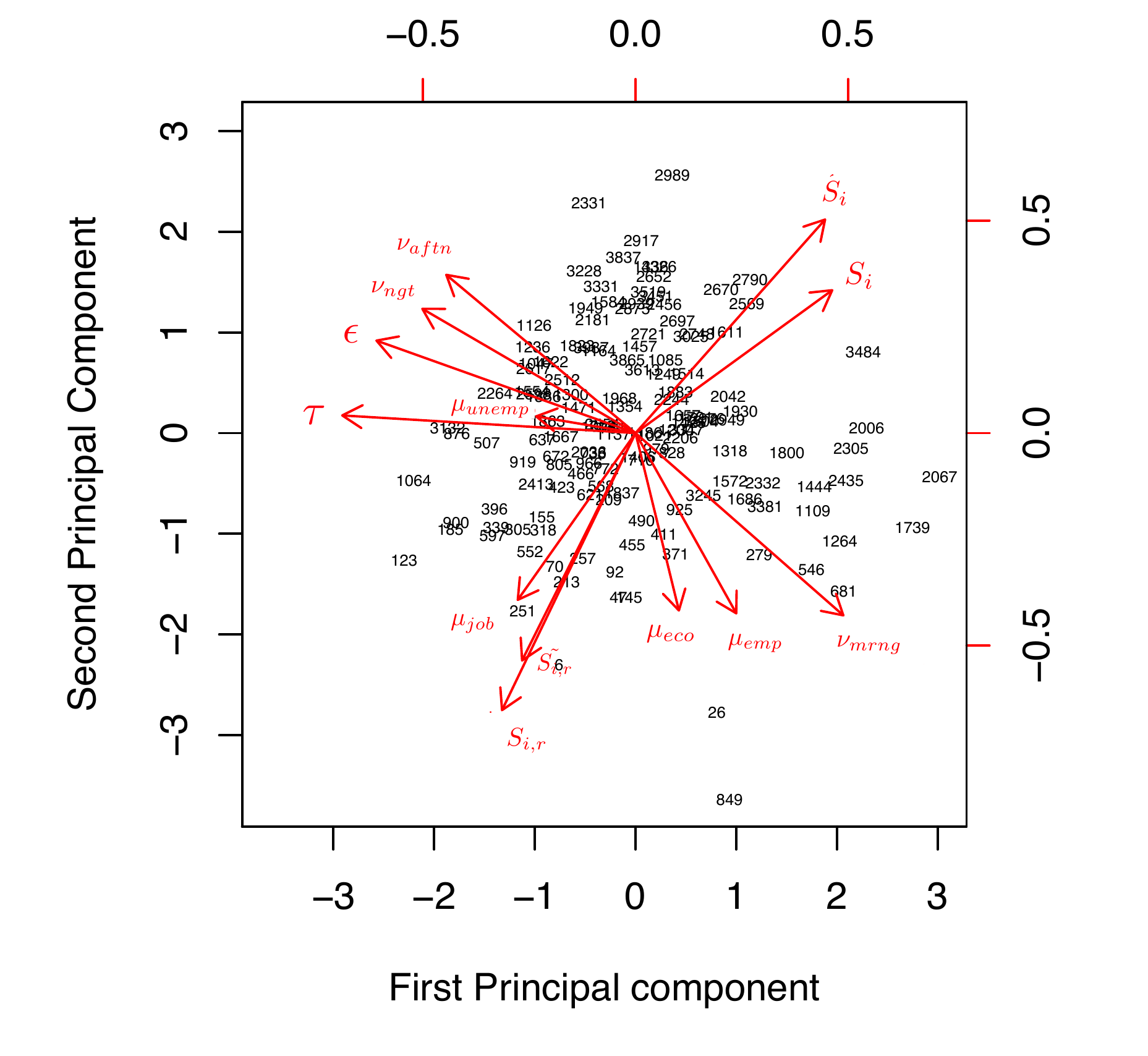}
	\caption{Top: Correlation matrix between the variables constructed from Twitter. Each entry in the matrix is depicted as a circle whose size is proportional to the correlation between variables and the sign is blue/red for positive/negative correlations. Blank entries correspond to statistically insignificant correlations with \%95 confidence. Bottom: Variables projection on the first two principal components given by PCA. We observe different groups of variables and collinearity between some of them.}
	\label{fig:correlation}
\end{figure}

Variables are constructed to reflect the behavior of areas in the different dimensions of Twitter penetration, social or geographical diversity, activity through the day and content. Correlation between variables does indeed show that variables within each dimensions hold strong correlations between them. As we can see in figure \ref{fig:correlation} social and geographical diversities are highly correlated between them, an expected fact given the {\em gravity law} accurate description of flows of people between geographical areas, but also the amount of communication between them. Same behavior is found for the group of variables in the activity group, while content variables are less correlated. Finally we find that both the penetration rate $\tau_i$ and fraction of misspellers $\varepsilon_i$ have a strong correlation with most of the variables. 

High correlation between variables might lead to collinearity effects \cite{SIwold1984collinearity} in the linear regression models, that is, some variables with predictive variable might have non-significant weights because they explain the same part of the variance. For instance, in Table \ref{table:coefficients} misspellers rate has a very strong predictive value but its $p$-value is too high to consider it significant. To test this hypothesis, we perform a principal component analysis (PCA) on the independent variables of the regression. Figure \ref{fig:correlation} exhibits the loadings of the different variables for the considered variables. The block structure showed in \ref{fig:correlation} results in similar directions of the variables in the first componentes of the PCA. We observe some groups of variables: on the one hand, geographical and social diversity seem to explain large part of the variance; on the other hand, we find a perpendicular group of variables formed by temporal activity; finally, penetration rate and misspellers fraction seem to represent a different independent direction of data, with high collinearity between them. This might explain the low statistical significance in the models of section \ref{sec:other areas}. In any case, the structure of the correlation matrix and the PCA results show that there is indeed information in all groups of variables and thus we have take a variable in each of them for our regression models.

\section*{S6. Misspellers detection}

In this work we will consider only tweets in Spanish, that is, since in Spain several languages live at the same time, depending on the part of the country, the first step is to reduce our Twitter dataset to those tweets that are written in Spanish. This task is carried out using the n-gram based text categorization R library {\it textcat} \cite{SIfeinerer2013textcat}.
Then, in order to decide whether a tweet has a misspelling or not, we need to establish some patterns to select from our set of tweets. Since we want to be sure that a detected mistake corresponds to a real misspeller, we will not consider the following cases:

\begin{itemize}
\item Lack of written accents. People tend to avoid writing accents when talking in a colloquial way.
\item Mistakes derived from removing {\it unnecessary} letters. The most common cases are removing a {\it h} at the beginning of a word (in Spanish the letter {\it h} is not pronounced), or replacing the letters {\it qu} by {\it k}. We understand that these mistakes can be motivated for the limitation of length in tweets, and not for a real misspelling.
\item In the same line, we neglect mistakes produced by removing letters in the middle of a word, whose pronunciation can be deduced without them.
\item We do not consider either mistakes related to features of specific areas in Spain. For example, in the south the pronunciation of {\it ce} and {\it se} is the same, what produces a big amount of mistakes when writing. However, since we want to extract objective and equitable conclusion over the whole Spanish geography, we neglect those misspellings that only appear in a specific area.
\end{itemize}

Likewise, we will consider as real misspellings the following mistakes:
\begin{itemize}
\item Adding letters. For example, writing a {\it h} at the beginning of a word that starts with a vowel.
\item Changing the special cases {\it mp}, {\it mb} by the wrong writings {\it np}, {\it nb}.
\item Mixing up {\it b} with {\it v}, {\it g} with {\it j}, {\it ll} with {\it y}, and {\it ex} with {\it es}. These are typical mistakes in Spanish, because they have the same, or a very close,  pronunciation.
\item Confusing the verb {\it haber} with the periphrasis {\it a ver}.
\item Separating a word into two ones, for instance, writing the word {\it conmigo} as {\it con migo}.
\end{itemize}
This way, our list of mispellings is composed of 617 common mistakes in Spanish, that cannot be attributed to the special features of Twitter or a specific region of Spain. Thus, one can expect that this selection provides an accurate and equitable method of detecting misspellers. Under these conditions, the number of users who wrote at least one misspelled word is 27055 (5.6\% over the whole population).

We analyze whether misspellers have different Twitter usage behavior from that people who do not make serious mistakes when publishing a tweet. Comparing the average number of tweets, it can be observed that misspellers tend to publish a larger number of tweets than those who did not made mistakes (144.71 against 23.72). This also emerges when the mean number of misspelling given the total number of tweets is considered. For users with less than approximately 30 published tweets in the observation period, the number of misspellings is almost zero whereas for users who publish more often, the mean number of misspellings scales sub-linearly with the number of tweets ($exponent \approx 0.33$).

\begin{figure}[t]
\centering
	\includegraphics[width=0.45\textwidth]{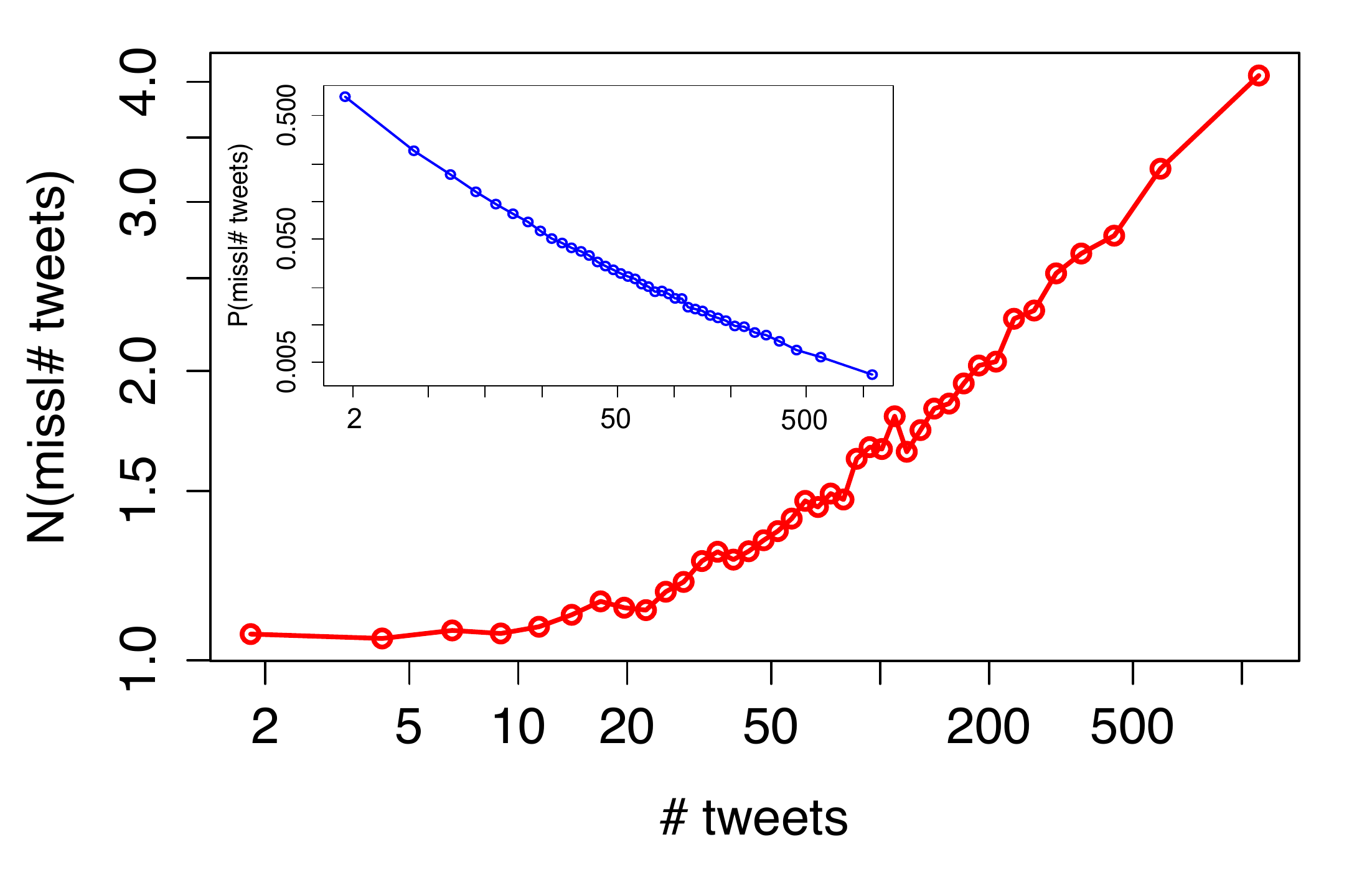}
	\caption{Number (red) and probability (blue) of observed misspellings given the number of tweets.}
	\label{fig:missprob_ntweets}
\end{figure}

Since we have observed a segmentation of Twitter population based on how accurate they write, we consider the misspeller rate as a proxy of the educational level of the cities. Large number of previous works in the literature have revealed the relationship between the economical status and the educational level of geographical areas and therefore it is natural to ask whether the observed misspellers rate is related to economy driven by the unemployment rate. To test this hypothesis, we consider cities populated with more than 5000 inhabitants to avoid subsampled cases. We find a strong positive correlation between the probability of finding a misspeller in a city and the unemployment rate (0.372, 0.491).

\section*{S7. Time window and unemployment}
\begin{figure}[t]
\centering
	\includegraphics[width=0.45\textwidth]{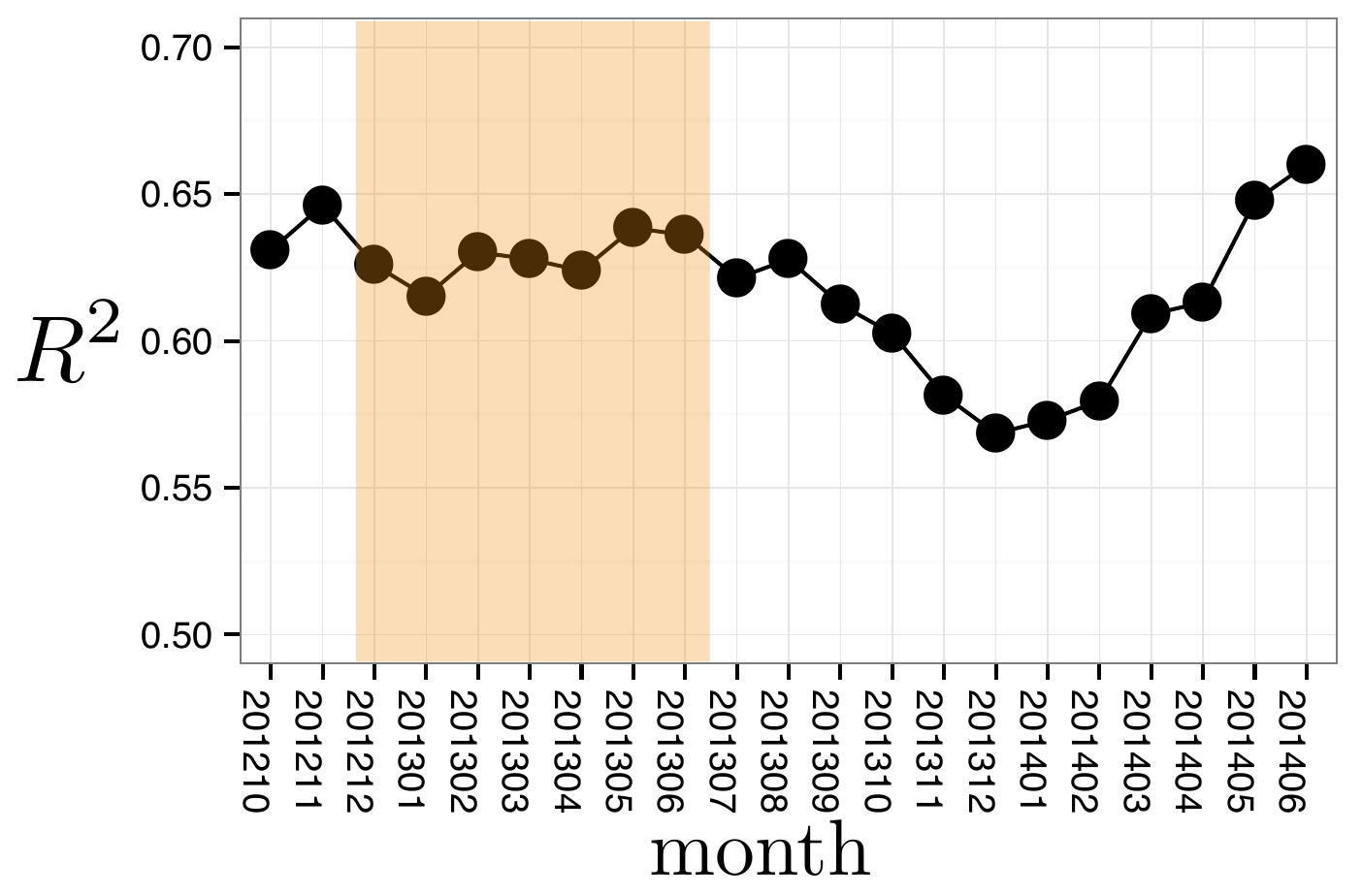}
	\caption{Explanatory power of the linear regression model when fitted against the unemployment data for different months. Gray (orange) area correspond to the time window in which Twitter data is collected and variables are constructed.}
	\label{fig:timewindow}
\end{figure}

In the definition of the variables we have aggregated the Twitter activity within a 7 months time window (from December 2012 to June 2013). Since unemployment has a significant variation along time, we investigate here what is the correlation and explanatory power of the Twitter variables for the values of unemployment determined at different months through the same time window in which Twitter data was collected. Or if the variables collected in that time window are more correlated with past or future values of unemployment. Figure \ref{fig:timewindow} shows the explanatory value of the model when the linear regression is done for values of unemployment of different months before, during and after the Twitter data time window. Although there is a small seasonal effect along the year, we see that the explanatory power remains around $R^2 = 0.6$, which suggest that our Twitter linear model retains its explanatory power even though unemployment changes considerably throughout the year. It is interesting to note that $R^2$ decays a little bit during the summer which means that our variables are less correlated with summer unemployment. Finally, unemployment used in the main article is from June 2013, i.e. the last month in the time window used to collect the data.

\section*{S8. Demographics does not explain unemployment}

Since unemployment rates are very large for the group of young people, a natural question is whether only demographic variables could explain the heterogeneity of young unemployment rates found in the geographical areas. To test this end we have built four linear models: the first one (named {\em Youth model} in Table \ref{table:coefficients}) is composed by the rate of young population as the only explaining variable; the second ones are  built based on only the Twitter variables considered in the main text (named {\em Twitter model (I)}) or just with those whose regression coefficients are statistically significant ({\em Twitter model (II)}); the third one is fitted with all the variables (named {\em All variables} model in Table \ref{table:coefficients}). In table \ref{table:coefficients} we show the summary of the regression for each model. Focusing on the explained variance by the model in terms of $R^2$, it can be checked that considering all Twitter variables is three times more explanatory than considering only the young people proportion. On the other hand, the comparison of $R^2$ for the {\em Twitter model} with the one for {\em All variables} and {\em Youth model} shows that the rate of young population does not provide a significant explanatory power. This semi-partial analysis shows that our Twitter variables retain a high explanatory power when the effect of young population rate is controlled.
\begin{table*}[t]
\begin{center}
\begin{tabular}{l c c c c c}
\hline
                  	& All variables	& Youth model 	& Twitter model (I) 	& Twitter model (II)\\
\hline
(Intercept)       	& $0.06$		& $-0.02$      	& $0.10^{***}$ 		& $0.09^{***}$	\\
 				  	& $(0.03)$		& $(0.03)$ 		& $(0.03)$ 			& $(0.027)$		\\
Young pop. rate   	& $0.66^{*}$	& $2.20^{***}$ 	&              		&		\\
 				    & $(0.30)$		& $(0.35)$ 		& 					&		\\
Penetration rate    & $8.20^{***}$	&            	& $8.57^{***}$ 		&$8.62^{***}$		\\
					& $(2.25)$		&			   	& $(2.22)$			&$(2.21)$		\\
Geographical diversity &$0.14^{***}$&        		& $0.15^{***}$ 		&$0.12^{***}$		\\
					& $(0.04)$		&				& $(0.04)$			&$(0.03)$		\\	
Social diversity 	& $-0.02$		&             	& $-0.03$      		&       	\\
					& $(0.02)$		&				& $(0.02)$			&		\\
Morning activity    & $-1.42^{***}$	&             	& $-1.30^{**}$ 		& $-1.28^{**}$	 	\\
					& $(0.41)$		&				& $(0.42)$			& $(0.41)$		\\
Misspellers rate    & $23.95$		&             	& $31.51^{*}$  		& $32.28^{*}$		\\
					& $(13.09)$		&				& $(12.78)$			& $(12.71)$		\\
\emph{Employment} mentions  & $0.34$ & 				& $3.17$			&    	\\
					& $(9.81)$		&				& $(9.86)$			&		\\ 
\hline
R$^2$             	& $0.65$			&$0.24$         	& $0.64$         		& $0.63$		\\
Adj. R$^2$        	& $0.63$			&$0.24$         	& $0.62$         		& $0.62$		\\
\hline
\multicolumn{4}{l}{\small{$^{***}p<0.001$, $^{**}p<0.01$, $^*p<0.05$}}
\end{tabular}
\caption{Regression table for the different statistical models. The {\em All variables} model includes both Twitter and rate of young population variables. {\em Twitter model (I)} includes only the variables described in the main article, while {\em Twitter model (II)} only includes those variables which are significant $p < 0.05$ in {\em Twitter model (I)}.}
\label{table:coefficients}
\end{center}
\end{table*}

\section*{S9. Unemployment models for other geographical areas}\label{sec:other areas}
While municipalities are very heterogeneous demographically, other administrative areas exist in Spain at large scales that could be used for our model of unemployment. As mentioned in section \ref{sec:community}, the smallest administrative division of Spain we have considered is that of the $~8200$  {\em municipalities}. At larger scales we have the 326 {\em counties} ({\em comarcas} in spanish) which are aggregations of municipalities. Finally, the largest geographical scale we considered is defined by 50 provinces ({\em provincias} in Spanish). In this section we compare the performance of our Twitter model for unemployment for the variables defined in those administrative areas and relate it to the geographical communities detected and used in the main paper (see section \ref{sec:community}). Not all the areas at different administrative divisions are considered in the model. To minimize the effect of areas in which the number of geo-tagged tweets is very small, we only consider the 1738 municipalities which have a Twitter population $\pi > 10$. Similarly, we only consider the 198 counties with $\pi > 100$. As we can see in Table \ref{table:modelareas} the model has a large explanatory power for areas equal or bigger than counties. As expected $R^2$ increases as the number of areas in the model is smaller, but the description level of the model is very low for provinces, for example. The best performance (high $R^2$ and high geographical description level) is attained at the level of the detected communities.

\begin{table*}[t]
\begin{center}
\begin{tabular}{l c c c c}
\hline
                  	& Communities 	& Municipalities &	Counties	 	& Provinces \\
\hline
(Intercept)       	& $0.10^{***}$ 		& $0.16^{***}$	&$0.11^{***}$	&$0.11^{*}$ \\
 				  	& $(0.03)$ 			& $(0.01)$		&$(0.03)$		&$(0.05)$ 	\\
Penetration rate    & $8.57^{***}$ 		&$4.01^{***}$	&$9.12^{***}$	&$10.47^{***}$\\
					& $(2.22)$			&$(0.59)$		&$(1.81)$		&$(1.97)$	\\	
Geographical diversity & $0.15^{***}$ 	&$0.02$			&$0.12^{***}$	&$0.08$		\\
					& $(0.04)$			&$(0.01)$		&$(0.03)$		&$(0.07)$	\\	
Social diversity 	& $-0.03$      		&$-0.01$	    &$-0.01$   		&$-0.03$	\\
					& $(0.02)$			&$(0.01)$		&$(0.02)$		&$0.07$		\\
Morning activity    & $-1.30^{**}$ 		& $-1.16^{***}$	&$-1.49^{***}$ 	&$-1.03$	\\
					& $(0.42)$			& $(0.14)$		&$(0.39)$		&$(0.88)$	\\
Misspellers rate    & $31.51^{*}$  		& $14.40^{***}$	&$14.09$		& 	\\
					& $(12.78)$			& $(2.51)$		&$(10.02)$		&	\\
\emph{Employment} mentions  & $3.17$	& $-0.71$   	&$2.41$			&$-3.17$\\
					& $(9.86)$			& $(0.89)$		&$(8.86)$		&$(12.29)$\\ 
\hline
Number of points	& $128$				& $1738$		&$198$			&$50$ \\
R$^2$             	& $0.64$         	& $0.22$		&$0.55$			&$0.65$\\
Adj. R$^2$        	& $0.62$         	& $0.21$		&$0.54$			&$0.61$\\
\hline
\multicolumn{4}{l}{\small{$^{***}p<0.001$, $^{**}p<0.01$, $^*p<0.05$}}
\end{tabular}
\caption{Regression table for the unemployment linear regression model in different levels of geographical areas. In the {\em Provinces} model, the misspellers rate has been removed from the model due to the large collinearity with the penetration rate.}
\label{table:modelareas}
\end{center}
\end{table*}

\section*{S10. Relative importance of the variables}
\begin{figure}[!t]
\centering
	\includegraphics[width=0.49\textwidth]{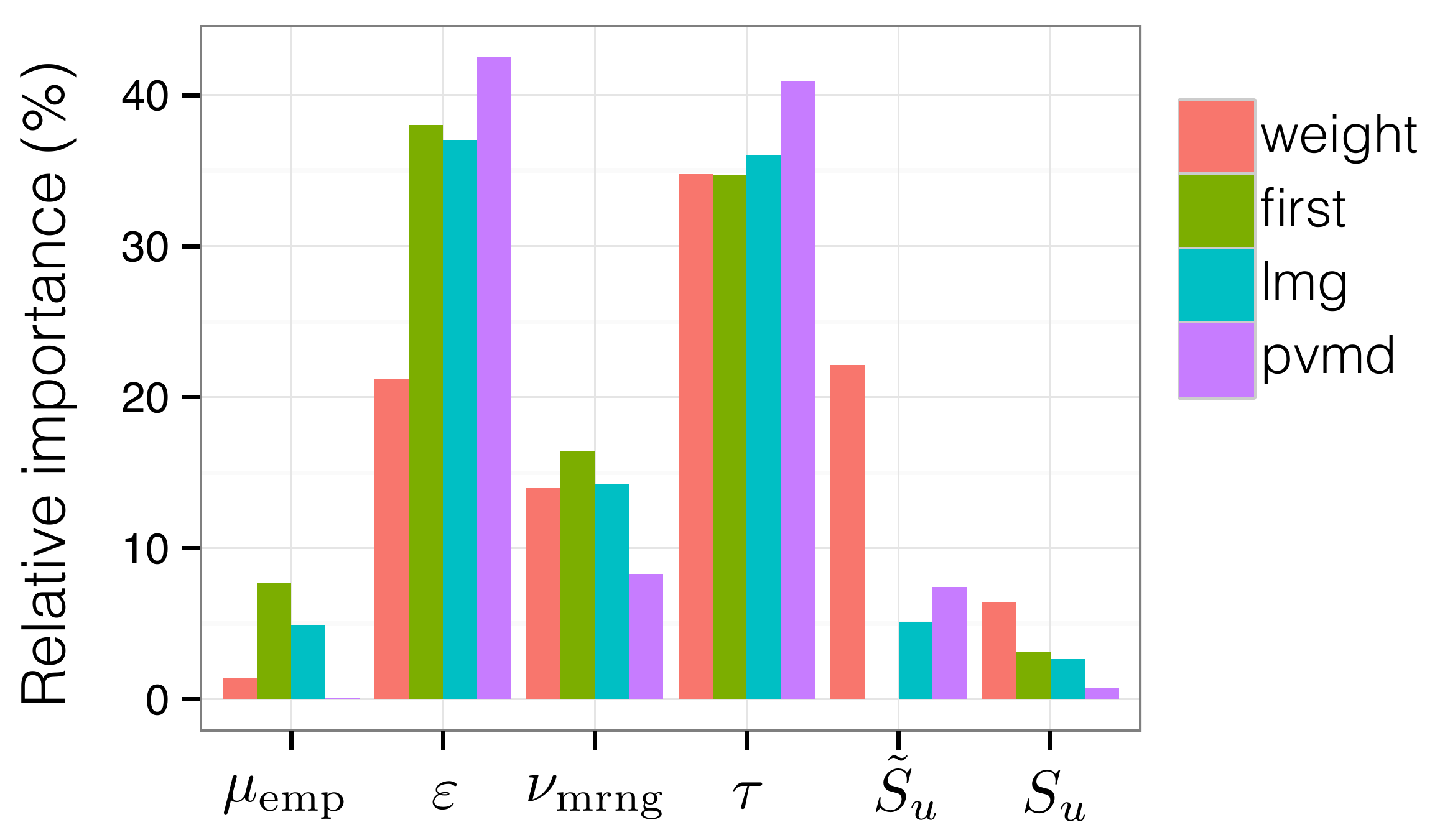}
	\caption{Relative importance of the variables (in percentage) in the unemployment model for different ways to calculate it.}
	\label{fig:relaimpo}
\end{figure}

To asses the relative importance of the variables in the unemployment model we have used several methods. They all give qualitatively the same results, with some variations for the statistically insignificant variables. Specifically, we have use
\begin{enumerate}
\item {\em (weight)}: Relative weight of the absolute values of the coefficients obtained in the linear regression when variables are scaled to have mean zero and variance one.
\item {\em (lmg)}: averaging over orderings, proposed by Lindeman, Merenda and Gold 
\item {\em (pmvd)}: The PMVD metric introduced by Feldman which an average over orderings as well, but with data-dependent weights 
\item {\em (first)}: The univariate $R^2$-values from regression models with one variable only.
\end{enumerate}
All these metrics are obtained using the {\it{relaimpo}} R package \cite{SIgromping2006relative}. The results for the young unemployment model are shown in figure \ref{fig:relaimpo} where we can see that different methods yield to similar relative importance of the variables, excepting perhaps for the diversity of mobility flows, a variable with a non-significant weight in the regression model.

\end{document}